\documentclass{aims}

\usepackage{txfonts,hyperref}
\def\typeofarticle{Research Article}
\def\currentvolume{}
\def\currentissue{x}
\def\currentyear{2023}

\def\ppages{xxx--xxx}
\def\DOI{}
\def\Received{November 2022}
\def\Revised{22 November 2022}
\def\Accepted{December 2022}
\def\Published{January 2023}

\numberwithin{equation}{section}

\begin{document}

\title{Population mobility, well-mixed clustering and disease spread: a look at COVID-19 Spread in the United States and preventive policy insights}  

\author{%
  David Lyver\affil{1,}, Mihai Nica\affil{1,}, Corentin Cot\affil{2,}, Giacomo Cacciapaglia\affil{3,}, Zahra Mohammadi\affil{1,}, Edward W. Thommes\affil{1,4,}, Monica-Gabriela Cojocaru\affil{1,}\corrauth
}

\shortauthors{the Author(s)}

\address{%
  \addr{\affilnum{1}}{Department of Mathematics, University of Guelph, 50 Stone Rd E, Guelph, ON N1G 2W1, Canada}
  \addr{\affilnum{2}}{Laboratoire de Physique des 2 Infinis Ir\`ene Joliot Curie (UMR 9012), CNRS/IN2P3, 15 Rue Georges Clemenceau, 91400 Orsay, France}
\addr{\affilnum{3}}{Institut de Physique des 2 Infinis de Lyon (UMR 5822), CNRS/IN2P3 et Universit\'e Claude Bernard Lyon 1, 4 rue Enrico Fermi, 69622 Villeurbanne Cedex, France}
  \addr{\affilnum{4}}{Sanofi, 1755 Steeles Ave W, North York, ON M2R 3T4, Canada}
  }

\corraddr{mcojocar@uoguelph.ca; Tel: 519-824-4120 x53295.
}

\begin{abstract}
The epidemiology of pandemics is classically viewed using geographical and political borders; however, these artificial divisions can result in a misunderstanding of the current epidemiological state within a given region. To improve upon current methods, we propose a clustering algorithm which is capable of recasting regions into well-mixed clusters such that they have a high level of interconnection while minimizing the external flow of the population towards other clusters. Moreover, we analyze and identify so-called {\it core clusters}, clusters that retain their features over time (temporally stable) and independent of the presence or absence of policy measures. In order to demonstrate the capabilities of this algorithm, we use US county-level cellular mobility data to divide the country into such clusters. Herein, we show a more granular spread of SARS-CoV-2 throughout the first weeks of the pandemic. Moreover, we are able to identify areas (groups of counties) that were experiencing above average levels of transmission within a state,  as well as pan-state areas (clusters overlapping more than one state) with very similar disease spread. Therefore, our method enables policymakers to make more informed decisions on the use of public health interventions within their jurisdiction, as well as guide collaboration with surrounding regions to benefit the general population in controlling the spread of communicable diseases.
\end{abstract}

\keywords{
\textbf{clustering, machine learning, SARS-CoV-2, epidemiology}
}

\maketitle

\section{Introduction and background overview}  

Pandemics caused by infectious diseases are becoming a constant threat in our globalized society. There are seasonal diseases like influenza, but also new diseases, often of zoonotic origin, like the recent case of COVID-19, cause by the coronavirus SARS-CoV-2. COVID-19 has become a pandemic starting in 2020 and it has left an important legacy in the form of extensive data covering various aspects relevant for the diffusion of the infection.
During the COVID-19 pandemic, traditional compartmental modeling of infections, based on ordinary differential equations (ODEs), has been employed very successfully in describing the evolution of the incidence of infections. More advanced versions of the traditional models have been proposed and tested, taking advantage of the unprecedented data availability. Data on people's mobility and behavior, on the adoption of public measures, on economic restrictions and public policy, were also fundamental in trying to make sense of the pandemic as it happened. As a hindsight exercise, one of the major questions currently on the topic of infectious disease spread is how best to use the available data in transmission models in such a way that new insights can be uncovered and new lessons/conclusions can be drawn from our common recent COVID-19 experience, should we be again faced with similar situations. 

Compartmental models are based on pioneering work in the early 20th century \cite{kermack1927contribution}. The first models where based on three compartments: Susceptible (S), Infectious (I) and Removed (R), leading to the famous SIR model. More recently, the compartment of Exposed (E) has been added to take into account the latency of the disease, leading to SEIR models. Such models are flexible enough to allow for interactions among separate geographical regions or age stratification.
In the former case, each region has its own set of SEIR-type ODEs with connecting terms to reflect importation of cases from other regions. From the administrative point of view, the regional division of population (for example counties in the USA, or public health regions in the province of Ontario, Canada) is based on various socio-demographic criteria, history, etc. Established literature in disease transmission has looked at regions as cities, with commuter traffic and/or regular travel between them \cite{arino2003,arino2004,thommes2016absenteeism}.

In general, SEIR-type models work well for large well-mixed and isolated populations within which the infections can easily spread \cite{anderson1979population}. This requirement is often not respected by administrative regional divisions. For instance, in the USA, counties are often sparsely populated or strongly connected to nearby counties by commuting population; states, instead, may encompass disconnected local areas or feature important cross-state commuting population. The main goal of our work is to recast small geographical units, like counties, in new well-mixed and isolated regions by use of available data on people's mobility. The new regions are defined by the following criteria: {\it minimizing mobility between the new regions} while creating {\it well-mixed sub-populations in each such new region}. Moreover, we introduce a notion of temporal stability of our clusters and use it to analyze the results throughout a 6 months time window.
Clustering of populations is not a novel concept; researchers have studied clustering populations in order to improve government legislation, re-imagine municipal infrastructure, and observe the environmental impacts of carbon emissions \cite{Hevesi2006, Laszlo2007}. While research has been active in this field, research into clustering regions based on interconnected mobility seems less represented in the literature, to the best of our knowledge.

Nevertheless, mobility networks have been used in conjunction with SEIR-type models in order to capture epidemiological dynamics of COVID-19 in urban populations \cite{chang2021,fields2021age,mohammadi2022human}. Some of the research was focused on individual city dynamics in order to capture the spread within these smaller regions \cite{chang2021}, while others have looked at quantifying the effect of mobility restrictions on the disease spread in Canada \cite{fields2021age} and in other countries around the world \cite{mohammadi2022human}. At larger scales, mobility can be used to understand the spread of the disease across continents: for instance, the case of the second wave in 2020 in Europe has been studied in \cite{cacciapaglia2020second} while the spread in the USA has been modeled at the census division level \cite{cacciapaglia2020better}. Many other cases have been studied in the literature that cannot be briefly summarized, in all cases struggling with the need of finding appropriate sub-population characterization.

For our purpose, we will employ a simple machine learning approach to define the new regions, based on the criteria and datasets mentioned above. There are three general approaches to classification problems: supervised, semi-supervised, and unsupervised  \cite{Fung:2001, Rodriguez:2019}. For the purposes of this article, we will focus on unsupervised classification, herein referred to as clustering \cite{Rodriguez:2019}.  In clustering, there is no information known regarding the true classification of data, unlike supervised and semi-supervised learning, wherein some or all of the information about the true classification is known \cite{Rodriguez:2019}. Clustering assigns classes to objects in a dataset \cite{Rodriguez:2019}.  Many different clustering algorithms exist and have a broad scope of applications, although no one clustering method is superior in all situations \cite{Rodriguez:2019}. 
Clustering algorithms vary in methodology and applications as described elsewhere \cite{Fung:2001,Xu:2015,Rodriguez:2019}. For the purposes of this research, our methodology most resembles a type of fuzzy clustering, with distinct differences. Fuzzy Theory clustering algorithms look to apply a probability to an object belonging to a cluster \cite{Xu:2015}. In applying a probability to the clustering, the membership of a data point is shared among all clusters and thus the boundaries of the clusters become fuzzy \cite{Fung:2001}. In general, these approaches aim to minimize a cost-function and achieve some local-minima \cite{Fung:2001}. Our algorithm looks to minimize over some cost-function and define the membership to each cluster as a probability. The difference is that our algorithm uses the maximum probability to define clusters after training.   

The structure of the paper is as follows: In Section \ref{sec2} we introduce the clustering algorithm adopted in this work; in Section \ref{sec3} we apply the algorithm to the USA counties in early 2020, focusing on the stability over the adoption of uneven measures; in Section \ref{sec4} we discuss the epidemiological implication of the adoption of the novel sub-populations; we finally offer our conclusions in Section \ref{sec5}.

\subsection{Present mobility data used} 

Part of the data that support these findings (U.S. COVID surveillance) are publicly available at \url{https://github.com/nytimes/covid-19-data}. The other part of the data (U.S contact rates) are available from data vendor Cuebiq  but restrictions apply to the availability of these data, which were used under license for the current study, and so are not publicly available. Data are however available from the authors upon reasonable request and with permission of Cuebiq. The Google mobility data and the Mobility Census data from Ontario are publicly available at \url{https://www.google.com/covid19/mobility/} and \url{https://www150.statcan.gc.ca/n1/en/catalogue/98-400-X2016391}, respectively.

\section{Clustering Algorithm} \label{sec2}

Gradient descent learning was first proposed by Louis--Augustin Cauchy in 1847 \cite{Cauchy1847} as an optimization algorithm suited to solving systems of coupled differential equations. Gradient descent algorithms minimize some objective function with respect to its variables \cite{Ruder2016}. This is done by calculating the gradient of the objective function and taking a small step in the opposite (decreasing) direction of the gradient. The step size is controlled by a learning rate, which is, in general, rather small. Recently, these algorithms have been used to optimize neural networks \cite{Ruder2016}. 

Here, we apply this class of algorithm to a network formed by small geographical units (i.e., counties) that are connected by people's mobility among them. The main goal is to define a set of macro-regions formed by clusters of the network nodes, which are maximally connected inside each cluster and minimally connected to other clusters.

\subsection{Basic strategy of the clustering algorithm}

We first define an arbitrary number of clusters, denoted by $N_\text{clusters}$, under the assumption that this number is much smaller than the number of nodes in the network, $N_\text{clusters}\ll N_\text{nodes}$. The main outcome of the algorithm will, therefore, yield the reorganization of the nodes into the desired number of clusters. 
We define a matrix of probabilities
    \begin{equation}
        P \in \mathbb{R}^{N_{\text{nodes}}\times N_{\text{clusters}}} \; ,
    \end{equation}
where each element $P_{ic} \in [0,1]$ denotes the probability that the node $i$ belongs to cluster $c$. Each row is bounded by conservation of probabilities to respect the following sum-rule:
    \begin{equation}
    \sum^{N_{\text{clusters}}}_{c=1} P_{ic} = 1\,.    \end{equation} 
The clustering algorithm must find the optimal probability matrix $P$ following a loss function, which is defined based on the desired properties of the clusters.
 
By using continuous probabilities instead of Boolean assignments, we can define a differentiable objective function that can be optimized with gradient descent: successive small improvements can be made to the algorithm assignments, rather than having abrupt changes when nodes are reassigned. We then deterministically assign each node to its maximum-probability cluster to get the best node-to-cluster assignment solution.

\subsection{Loss functions}

The optimization of the node-to-cluster assignment is based on a loss function that depends on the probability matrix $P$. It measures how accurately any value of $P$ matches the required features of the clusters. 
This loss function should evaluate to a large value when we have an inaccurate solution, but close to zero when we have an accurate node-to-cluster assignment solution. 

To fulfill our purposes, the loss function must have two parts, as we need to regulate two important measures:  {\it low mobility interactions among clusters} and  {\it low population difference among clusters}. Henceforth, we define the loss function as a convex combination of two terms:
\begin{equation}
\label{totalloss}
\text{Loss}_{\text{Total}}=\alpha_{\text{Int}}\  \text{Loss}_{\text{Int}} + \alpha_{\text{Pop}} \ \text{Loss}_{\text{Pop}}\,,
\end{equation}
where the constants $\alpha_{\text{Int}} , \alpha_{\text{Pop}} \in \mathbb{R}$ control the relative strength of the two requirements. Once the convex weights are fixed, we employ a gradient descent method to minimize the loss function and find the optimal probability matrix $P$.

The {\it low mobility interaction} $\text{Loss}_{\text{Int}}$ takes into account the interactions between clusters, measured in terms of the population mobility among the nodes of the network. Hence, this measure relies on mobility data, expressed in terms of an interaction matrix, $\text{Interaction}_{ij}$, whose elements are proportional to people's flow from node $i$ to node $j$. The loss function sums the interactions among all nodes belonging to different clusters, and it is defined as follows:
\begin{equation} \label{eq:LossInt}
\text{Loss}_{\text{Int}}:=\sum_{i,j=1}^{N_{\text{nodes}}}\text{Interaction}_{ij}\ \mathbb{P}_\text{different}(i,j) 
\end{equation}  
with $\mathbb{P}_\text{different}(i,j)$ being the probability of node $i$ to be in a different cluster than node $j$.
By the definition of the probability matrix $P$, we have
\begin{equation}
    \mathbb{P}_\text{different}(i,j) = 1 - [P\cdot P^T]_{ij} \; ,
\end{equation}
hence the loss function can be written in terms of matrix operations as:
\begin{equation}
    \text{Loss}_{\text{Int}}=\text{Tr}(\text{Interaction} \cdot (\mathbf{1} - P\cdot P^T)) \,,
\end{equation}
with $\mathbf{1}$ being defined as a matrix of ones.

The {\it low population difference}  $\text{Loss}_{\text{Pop}}$ forces the solution to contain clusters of approximately equal population. The main purpose of this term is to force the algorithm away from a trivial solution where all nodes are joined in a single giant cluster while the other clusters are left empty, which trivially minimizes the inter-cluster interactions. We define the loss function as follows:
\begin{equation}
    \text{Loss}_{\text{Pop}}:=\sum_{c=1}^{N_{\text{clusters}}}\left(\mathbb{E}_P[\text{Population of cluster } c ] - \frac{\text{TotalPop}}{N_{\text{clusters}}}\right)^2 \,,
\end{equation}
where $\mathbb{E}_P$ denotes the  population of cluster $c$ based on the node assignment given by the probabilities $P_{ic}$ and TotalPop is the total population in the network. 
Defining a vector of the node populations $\text{Pop}_i$, we have
\begin{equation}
    \mathbb{E}_P[\text{Population of cluster } c]=\sum_{i=1}^{N_{\text{nodes}}}P_{ic}\ \text{Pop}_i=(P^T\cdot \text{Pop})_c\,.\end{equation}
Hence, the loss function can be written in terms of matrix operations as:
\begin{equation} \label{eq:softmax}
    \text{Loss}_{\text{Pop}}=\left|P^T\cdot\text{Pop} - \frac{\text{TotalPop}}{N_{\text{clusters}}}\right|^2\,.    
\end{equation}

\subsection{Gradient descend algorithm}

There is a subtle aspect in the implementation of a gradient descent algorithm to our problem. In fact, as the variables $P$ must respect $P_{ic} \in [0,1]$ and $\sum^{N_{\text{clusters}}}_{c=1}P_{ic} = 1$, the gradient descent is not ideal as it would be applied to a constrained optimization problem with both box and linear constraints. Hence, to simplify the implementation, we apply the algorithm to a parametric matrix $X \in \mathbb{R}^{N_{\text{nodes}}\times N_{\text{cluster}}}$, where each entry $X_{ic} \in [-\infty, \infty]$ is unconstrained, and obtained P from $X$ via the {\it softmax} function:
\begin{equation}
P_{ic}=\frac{e^{X_{ic}}}{\Sigma^{N_{\text{clusters}}}_{c^\prime=1}e^{X_{ic^\prime}}}\,.
\end{equation}
This is a common implementation artifact used in machine learning.
To find a good cluster assignment configuration, the parameters at every step were updated using automatic differentiation in the grad package from the {\it Jax} library in {\it Python} using the following loss function:
\begin{equation}
    X_{\text{new}}=X_{\text{old}}-\text{StepSize}\cdot\nabla_X \text{Loss}(X)\,.
\end{equation}

\section{Implementation Details for USA data} \label{sec3}

We apply the clustering algorithm to a network made of 3,102 counties and county equivalents, located within the 50 states and the District of Columbia (DC). In total, there exist 3,144 counties and county equivalents within the USA, however 42 counties were missing from the Cuebiq dataset, and they are not included within our network. These missing counties will appear in white on the maps of the USA, as seen for instance in Figure~\ref{fig1}. 

The mobility data was provided from Cuebiq and it contains the number of users of their proprietary app traveling from county $i$ to county $j$ normalized by the number of users seen in county $i$. For each county, the data includes the 15 largest flows to other counties on a weekly timescale, hence the entries are largely dominated by commuter travelers among nearby counties. Airborne travelers, while not explicitly excluded, are numerically smaller than commuter and ground based ones and often do not make it above the cut or remain subleading. We consider the data from Cuebiq users as a good proxy of the total population of each county, as confirmed by the provider.
Using this data from Cuebiq we were able to construct the 3,102 by 3,102 flow matrix used in Eq.~\eqref{eq:LossInt}, where the matrix entries represented the flow from county $i$ to county $j$ normalized by the population of the county of origins. The population data was obtained from \url{https://raw.githubusercontent.com/Zoooook/CoronavirusTimelapse/master/static/population.json}.

\subsection{Initialization}

In order to speed up the convergence of the algorithm, the variable $X$ matrix was initialized using physical proximity among counties. To do this, we generated $N_{\text{clusters}}$ ``initialization central points'' (ICPs) spread across the USA, and then initialized the $X$-values in the algorithm for the $N_{\text{counties}}$ nodes proportionally to the distances from these points. The distances are computed from the geographical center of each county $i$ and the ICPs. Specifically, we set:
\begin{equation}
X_{ic} = -0.1 \cdot \text{distance}( \text{Center of County }i, \text{ICP }c)\,,
\end{equation}
where  $1\leq i \leq N_{\text{counties}}$ is the county index and $1\leq c\leq N_{\text{cluster}}$ is the cluster index.

\begin{figure}[H]
\begin{centering}
\includegraphics[scale=0.20]{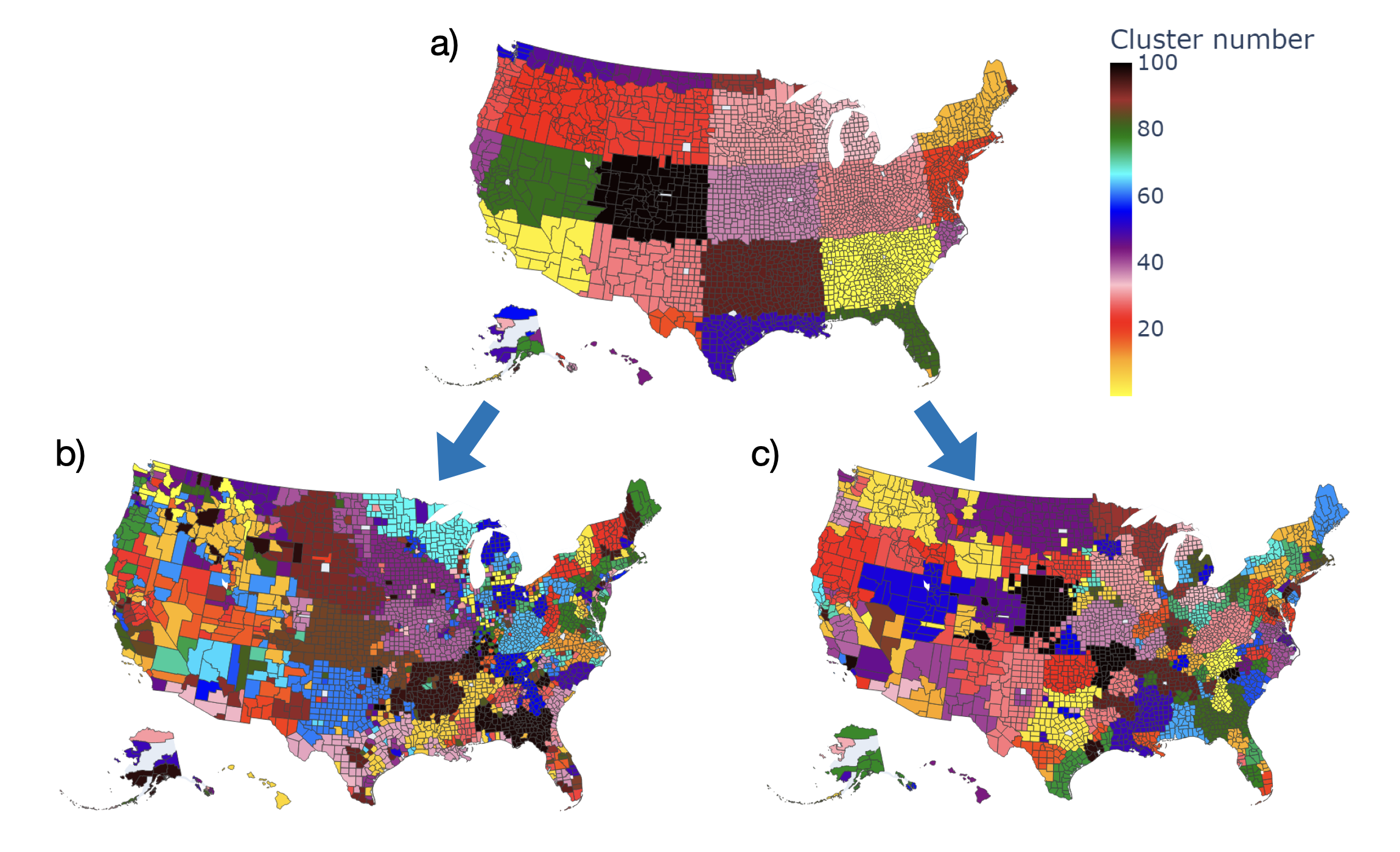}
\caption{Initialization based on geographic vicinity for 100 clusters (panel a). The outcome of the clustering algorithm after 50,000 gradient descent steps is shown for: (b) $\alpha_\text{Int} = \alpha_\text{Pop} = 0.5$ and (c) $\alpha_\text{Int} = 0.99$, $\alpha_\text{Pop} = 0.01$. The color gradient has no meaning other than the identification of each cluster.} 
\label{fig1}
\end{centering}
\end{figure}

In this way, at initialization, the algorithm assigns the highest probability of belonging to the cluster of the nearest ICP. In practice, the ICPs are defined as a rectangular grid of equally spaced points covering the USA (including Alaska and Hawaii), see Figure \ref{fig1}a for an example with $N_\text{clusters} = 100$. During the initialization process only a subset of clusters are populated, the remaining are later populated by the algorithm. For instance, the clusterings in Figures \ref{fig1}b and \ref{fig1}c were obtained for two different loss function combinations and after $50,000$ gradient descent steps.
As it can be seen in Figure \ref{fig1}b, sizable values of $\alpha_\text{Pop}$ force the clusters to have similar population, however creating heterogeneity in the their geographical extension and discontinuities. This is due to the very heterogeneous distribution of the population, which leads to densely populated counties and very sparsely populated ones. As a result, the clustering in Figure \ref{fig1}a results in clusters consisting of a handful of counties near cities, in contrast to very extended and discontinuous ones in more rural areas. To adjust the outcome, we reduced the impact of the population requirement by choosing $\alpha_\text{Pop} = 0.1$ and $\alpha_\text{Int} = 0.99$. This resulted in the clustering in Figure \ref{fig1}b, leading to more comparable and geographically continuous clusters. We then tested the algorithm with various numbers of clusters, obtaining comparable results.

\begin{figure}[H]
\begin{centering}
\includegraphics[scale=0.21]{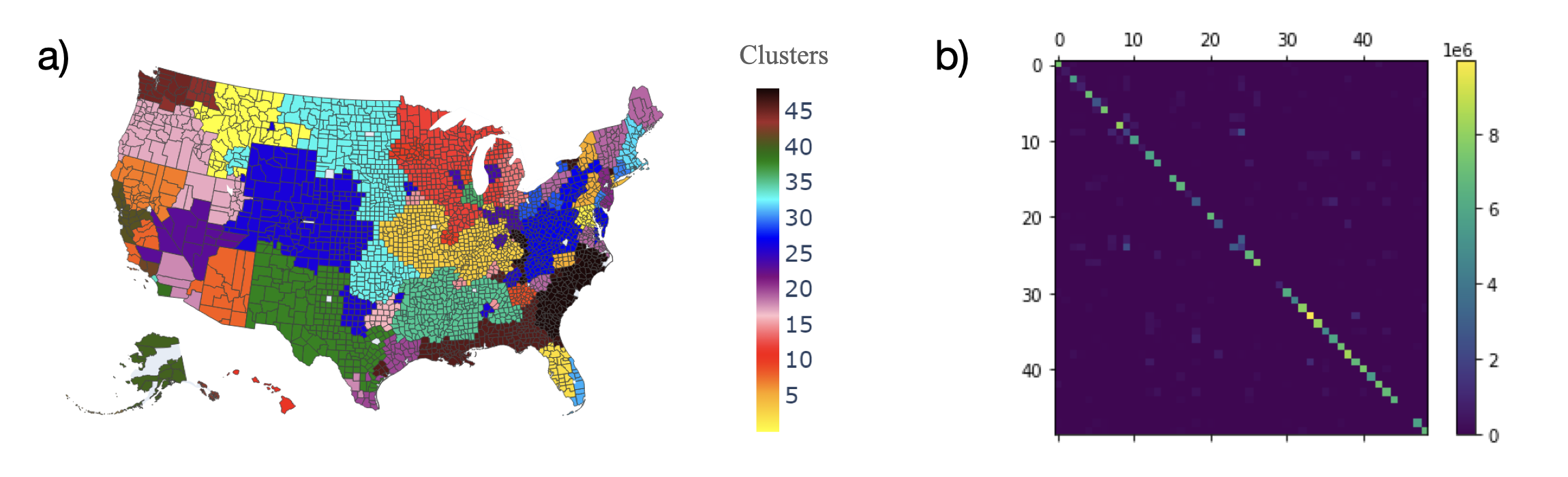}
\caption{Outcome of the clustering for pre-pandemic mobility data (first week of January 2020). The 49 clusters are shown in the left panel (a), while the mobility levels among clusters is represented in the heat map on the right panel (b). Diagonal entries show the mobility among counties inside the same cluster, while off-diagonal entries indicate the level of mobility inter-cluster.}
\label{fig4}
\end{centering}
\end{figure}

In the remainder of this work we will focus on $N_\text{clusters} = 49$, which provides a number of clusters closely comparable to the number of states (50 + DC). Hence, we obtained the 49 clusters with pre-pandemic mobility levels, where the data was taken from the first week of January 2020.  As a working point, we used skewed weights $\alpha_\text{Int}=0.99$ and $\alpha_\text{Pop}=0.01$ to minimize cluster-to-cluster mobility flows. The obtained clusters are visualized in Figure \ref{fig4}a. 
To test the residual level of mobility inter-cluster, we computed the following matrix at the end of the algorithm run
\begin{equation}
    M_\text{inter-cluster} = X^T_\text{out} \cdot \text{Interaction} \cdot X_\text{out}\,,
\end{equation}
where $X_\text{out}$ is the final value of the variable matrix $X$ used to define the cluster probabilities via Eq.~\eqref{eq:softmax}. The entries of this matrix are visualized in Figure \ref{fig4}b.

 The vast majority of the activity is detected on the diagonal, which represent the mobility among counties belonging to the same clusters. Instead, off-diagonal entries feature very small values, indicating very limited mobility among clusters belonging to different clusters. This result, therefore, validates the effectiveness of the clustering algorithm.
\subsection{Cluster stability over time during the COVID19 pandemic} 

Over the course of a pandemic, people's mobility continually changes as different regions are imposed various non-pharmaceutical measures in order to contrast the spread of the disease. During the early phases of COVID19 we have seen some states/regions being placed in lockdown, while others had lighter restrictions. In principle, such changes could affect how counties are clustered by our algorithm via changes in the interaction matrix.

To test the stability of the algorithm results, we considered the first six months of 2020, which saw the first diffusion of the infections and the most severe travel and local mobility restrictions. Hence, we  constructed six different clusterings using the mobility data from the first week of each month from January to June, 2020. We then use the output probability matrices $P_\alpha$, where $\alpha$ labels the month, to check the stability of the output.

\begin{figure}[H]
\begin{centering}
\includegraphics[scale=0.45]{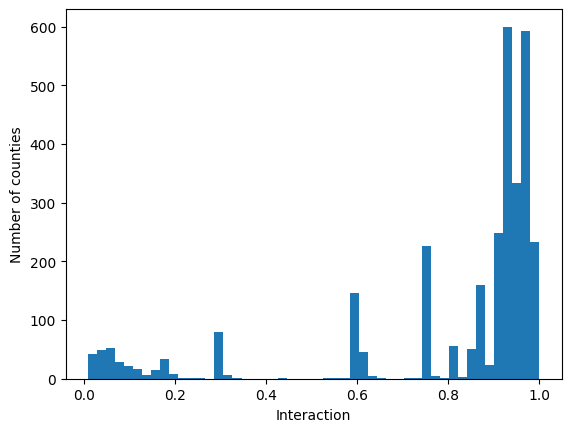}
\includegraphics[scale=0.45]{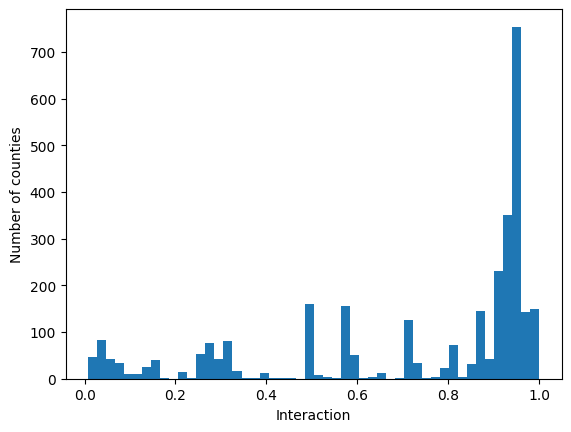}
\includegraphics[scale=0.45]{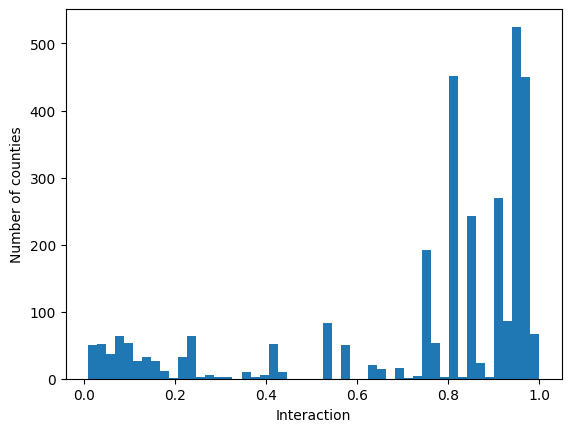}
\includegraphics[scale=0.45]{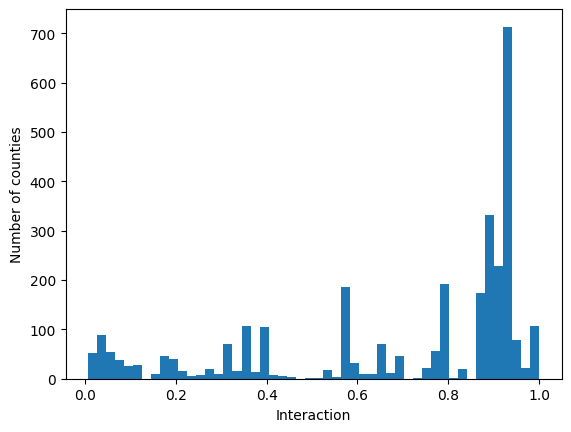}
\includegraphics[scale=0.45]{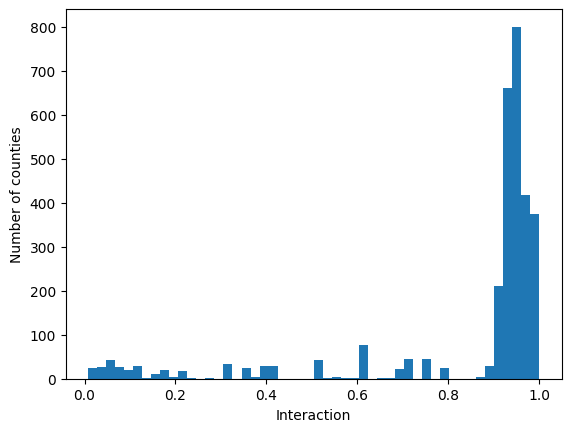}
\caption{County interaction across all six clusterings from: January to February, February to March, March to April, April to May, May to June }
\label{fig11}
\end{centering}
\end{figure}

To do so, we first define a product matrix $M_\alpha$ which traces over the clusters:
\begin{equation}\label{14}
M_\alpha = P_\alpha \cdot P^T_\alpha\,.
\end{equation}
Each entry $M_{ij} = \sum_{c=1}^{N_\text{clusters}} P_{ic} P_{jc}$ measures how likely the county $i$ belongs to the same cluster as the county $j$, as the product of the two probability rows is maximized if the two coincide. To express this similarity measure more objectively, we normalize the entries of the matrix in Eq.~\eqref{14} as follows:
\begin{equation} \label{eq:normM}
    \Bar{M}_{ij} = \frac{M_{ij}}{\sqrt{\sum_{j=1}^{N_\text{nodes}} M_{ij}^2}},\;\; \text{for}\;\; i \in [1,N_\text{nodes}]\,.
\end{equation}
The similarity measures are expressed by the diagonal entries of the product matrix
\begin{equation} \label{eq:S}
    S_{\alpha \beta} = \Bar{M}_\alpha \cdot \Bar{M}^T_\beta\,.
\end{equation}
In fact, for $\alpha = \beta$, the diagonal entries are all equal to 1 thanks to Eq.~\eqref{eq:normM}. For $\alpha \neq \beta$, closeness to unity for the diagonal entries measures how similar the two clusterings $\alpha$ and $\beta$ are to each other. We then construct the similarity matrix $S$ for clusterings stemming from consecutive months from January to June, 2020. The distribution of the diagonal entries for the five cases are shown in Figure~\ref{fig11}.

\begin{figure}[H]
\begin{centering}
\includegraphics[scale=0.18]{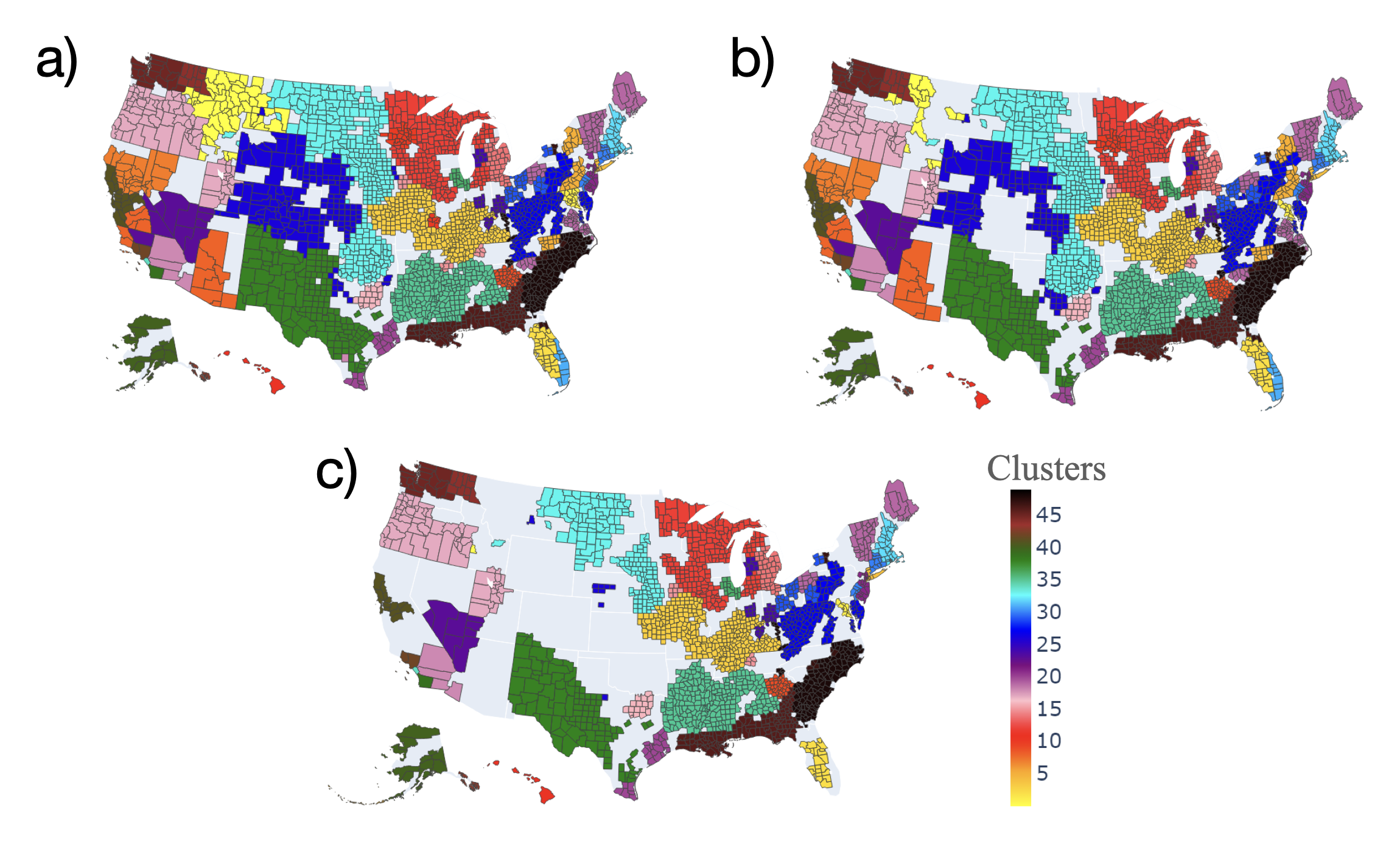}
\caption{Pruned unstable clusters in white for a threshold of: (a) $0.1$, (b) $0.25$ and (c) $0.5$. The remaining counties are associated to their most likely cluster over the 6-month period. Only 43 core clusters remain populated, while 6 are empty. }
\label{fig14}
\end{centering}
\end{figure}

These results show that the majority of counties remain in the same cluster over the 6-month period, as the majority of the entries remain very close to 1. This fact allows us to distinguish counties that consistently belong to the same cluster, and counties that flip between different counties (at least once). The pruning of unstable counties can be performed by keeping only clusters whose diagonal $S$-entries remain above a given threshold over the 6-month period under study. We show in Figure~\ref{fig14} the result for three values of the threshold: $0.1$, $0.2$ and $0.5$. The maps clearly show that inconsistent clusters tend to be located within less populated areas, while densely populated areas remain stable.

As an interesting highlight, pruning with threshold of 0.5 seems to agree quite well with the rural-urban map of the U.S. which we include in Figure \ref{fig:ruralUS} right panel, as presented in \cite{golding2020tracking}:
\begin{figure}[h]
\centering
\includegraphics[scale=0.2]{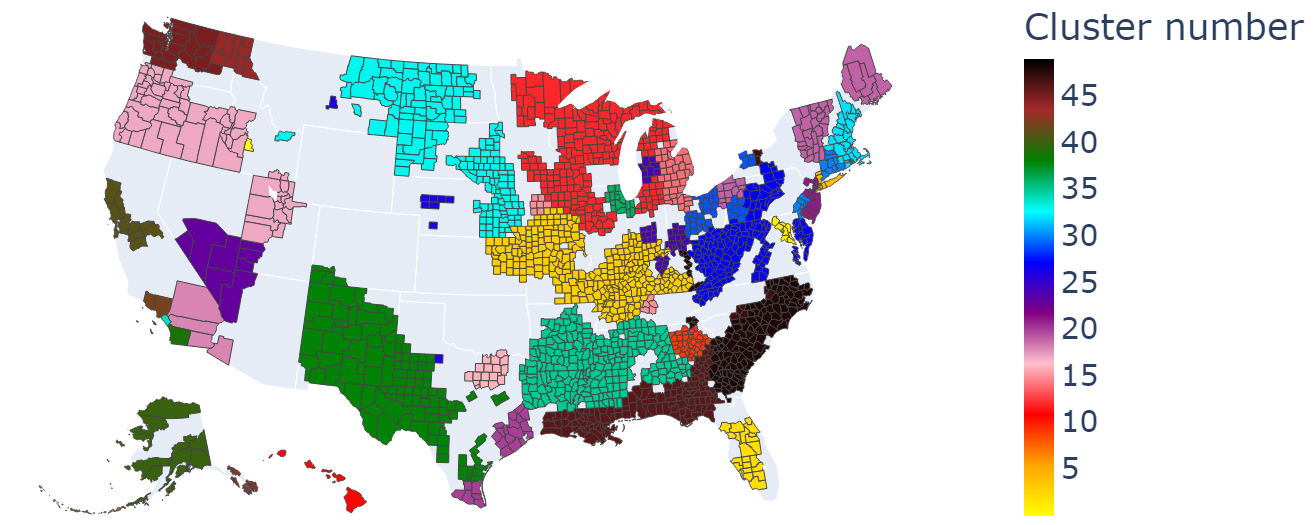}
\includegraphics[scale=0.2]{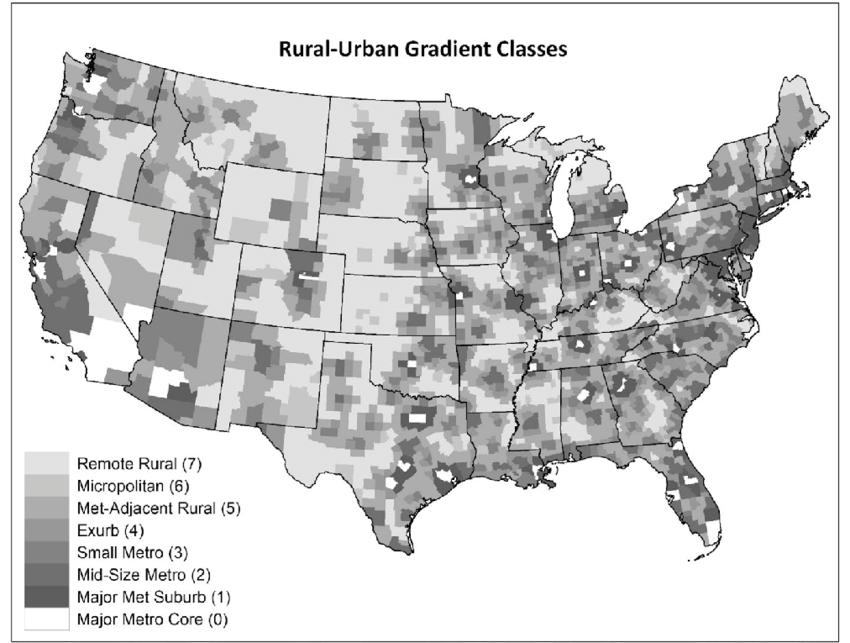}
\caption{Core Clusters of the USA with an interaction threshold of 0.5 (left) and Rural-urban map of U.S.. We see that a threshold of 0.5 in our core cluster definition regains the population density divide of rural vs. urban areas very well - both left and right figures depict rural areas in lightest shades.}
\label{fig:ruralUS}
\end{figure}
We find a good compromise to define stable clusters after pruning counties with a threshold of $0.25$, as shown in Figure~\ref{fig14}b, because we are not looking to exclude all rural counties from our analysis. Hence, a pruning with a threshold of 0.25 allows us to define 43 core clusters to be used for further analyses in the following section (6 clusters being emptied). 
We also note that we can check the effect of the population constraints we imposed in the clustering algorithm. In Figure~\ref{fig16} we compare the distribution of population in the 50+DC states and in the 43 core clusters remaining from above pruning. One can see that most core clusters have a population around 8 millions, while state populations are concentrated around smaller values with a few very populous exceptions. This shows that even the small weight $\alpha_\text{Pop} = 0.1$ is sufficient to obtain clusters with fairly balanced populations.
\begin{figure}[H]
\begin{centering}
\includegraphics[scale=0.5]{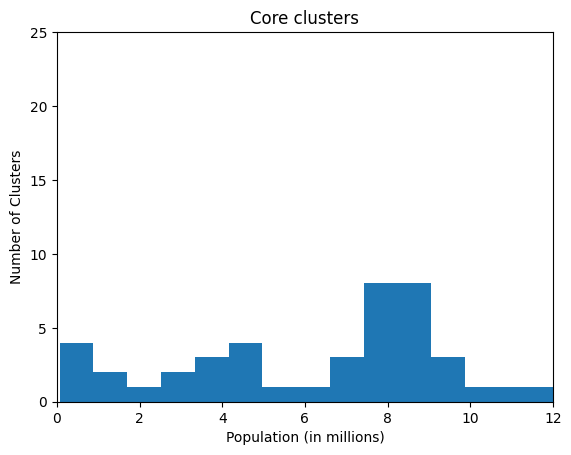}
\includegraphics[scale=0.5]{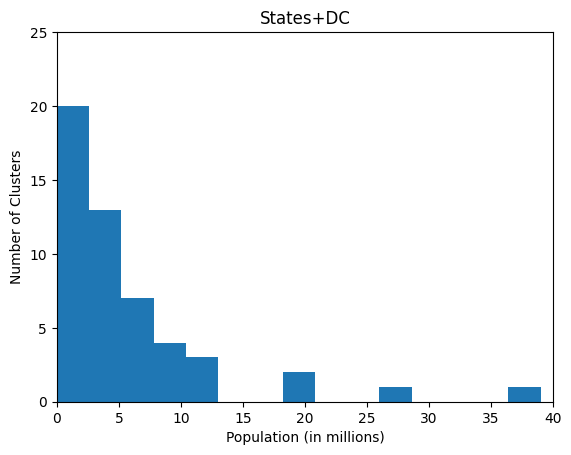}
\caption{Population distribution for Core Clusters (left) and States+DC (right).}
\label{fig16}
\end{centering}
\end{figure}

\section{Analysis of epidemiological data from a clustering perspective} \label{sec4}

COVID-19 incidence data is available at the county level in the USA. However, the small populations of many counties provide statistically poor data, which do not allow to clearly identify growing patterns. Hence, it is important to aggregate the data into larger geographical units.
By applying the clustering algorithm we aimed at identifying well-mixed regional sub-populations. Having achieved our goal mathematically, we now focus our attention on the analysis of the epidemiological data in the newly formed sub-populations in the core clusters. As compared to a state-level analysis, our approach permits a new view on the initial COVID-19 spread, highlighting indirectly the importance of flights and case importations. As a consequence, our results call for more localized preventive measures in large states (such as California), and the need for coordination/cooperation among neighboring states in staving off disease spread.

 \subsection{Disease spread - core cluster view}
 
Starting from incidence data provided by the New York Times at the county level, we aggregated the data both at state and core cluster level. Our aim is to study the growth of the number of infections and compare different regions. 
We assumed that, at the beginning of a wave of infections, the incidence number $\text{inc}(t)$ is given by an exponential curve of the type:
\begin{equation}
\text{inc}(t)=\text{inc}(0)\ e^{\rho t}\,.
\end{equation}
This behavior would signal the phase of exponential growth in the infections and the start of an epidemiological wave. 
Hence, we can compute a time-series of the exponential growth factor:
\begin{equation}
\rho =\ln \frac{\text{inc}(t+1)}{\text{inc}(t)}\,,\;\; \text{ with }\;\; \text{inc}(t)\neq 0\,.
\end{equation}
The growth factor $\rho$ is computed from the initial phase of nearly exponential growth in the neighborhood of the disease-free equilibrium state, corresponding to a phase of linear growth in $\log(\text{inc})$ with slope $\rho$. Using the growth factor estimates as above and the closed form formula for the initial reproduction number $R_0$ based on $\rho$ (see for details the work of \cite{ma2020estimating}\footnote{In a typical SEIR model, where $\sigma$ is the susceptible to exposed rate and $\gamma$ is the recovery rate, then $R_0=\frac{(\rho+\sigma)(\rho+\gamma)}{\sigma\gamma}$, where $\rho$ is the exponential growth factor.}) 
, we identify the initial fastest phase of nearly unchecked growth in any given region with the help of a piecewise linear fit to the log of the incidence.  We utilize the \texttt{R} function \texttt{dpseg()}, which is a part of the \texttt{dpseg} package, \url{https://cran.r-project.org/web/packages/dpseg/index.html}.  
This function uses a dynamic programming algorithm to generate an optimal piecewise linear fit to a time series, which balances goodness of fit against an (adjustable) penalty for each additional segment. We then identified the earliest segment with the steepest positive slope (largest $\rho$) as corresponding to the initial near-unchecked exponential growth phase.

\begin{figure}[H]
\begin{centering}
\includegraphics[scale=0.21]{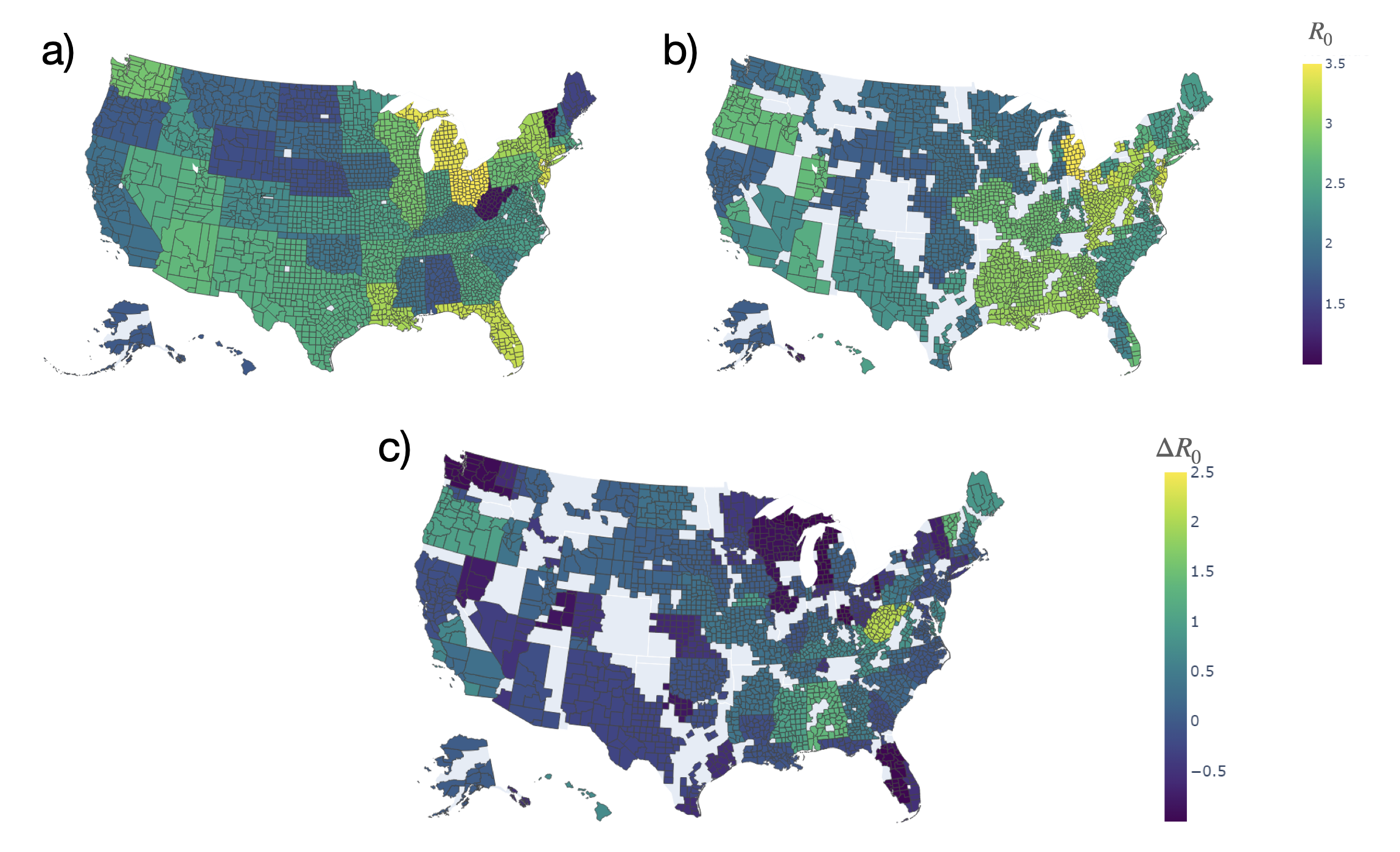}
\caption{$R_0$ values across the USA counties computed via State-level aggregation (a) and core cluster aggregation (b). We see that there are some differences in $R_0$ values on the clustering level, specifically in the North-East part of the USA. Panel (c) shows the difference between core cluster-level and state-level $R_0$ values across the USA counties belonging to core clusters. Areas in yellow highlight where a core-bases analysis would provide a much larger $R_0$ than a state-level analysis would suggest. The most prominent cases are counties in West Virgina, Vermont and Oregon. The white counties are either missing from the dataset or do not belong to any core cluster.}
\label{fig18}
\end{centering}
\end{figure}

In Figure~\ref{fig18} we visualize the values of $R_0$ obtained at state level (a) and by use of the core clusters (b). Here, the initial reproduction number $R_0$ is computed over week 9 to week 14 of 2020, that is the period from February 24th to March 16th, 2020. The results are fairly compatible, however a visual comparison of the two maps show that some counties are characterized by very different values.

To better visualize the difference between the two approaches - state vs. core clusters - for each county we computed the difference in the local $R_0$ obtained by the two different aggregations:
\begin{equation}
    \Delta R_0 = R_0 (\text{core cluster}) - R_0 (\text{state})\,,
\end{equation}
and show the values in Figure~\ref{fig18}c, where the color gradient corresponds to the size of the difference.
We see that areas in yellow highlight the biggest differences, meaning that the $R_0$ values in those clusters were higher than the state-case data $R_0$ indicated. The darker blue areas indicate that the state-case data gave a higher value of $R_0$ than the clustering data. From a policy perspective, the yellow areas are the important ones, as their presence means that, looking strictly at the geographic state level, policymakers may feel optimistic about their state-wide $R_0$ values when, in effect, due to people's mobility, the initial force of infection is much higher in many of their counties (see Figure \ref{fig18}c). The core cluster analysis, therefore, provides more reliable results for local communities living in a fraction of a state counties, where localized measures could be implemented to limit the incidence of the disease.

\subsection{Geographical spread of the infection in the USA}

\begin{figure}[H]
\begin{centering}
\includegraphics[scale=0.22]{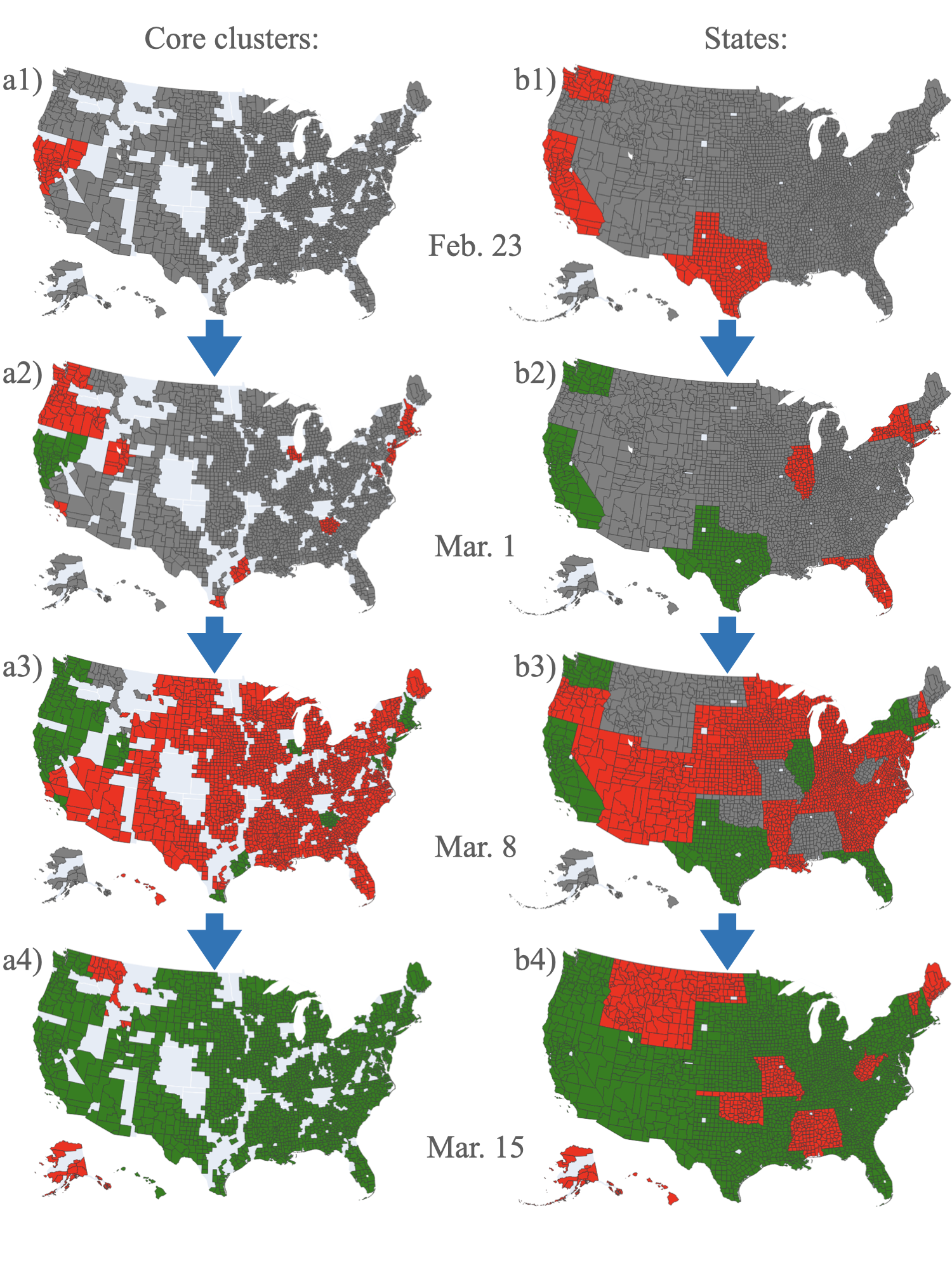}
\caption{Weekly time-lapse of the initial spread of COVID-19 from February 24 until the March 15, 2020. The left panels depict the initial spread in 43 core clusters, while the right panels depict the initial spread in USA states. Territories are highlighted in red during the week when they transition from disease-free equilibrium to the first exponential growth of the incidence number. Green indicates regions where the above transition has already occurred.  White indicates counties that have been pruned from the core clusters (left panels only) or absent in the dataset.}
\label{fig19}
\end{centering}
\end{figure}

The algorithmic definition of core clusters also provides a new view on the initial geographical spread of COVID-19 cases within the USA. In particular, it allows to see how the disease spreads geographically to the most populated areas, which are also served by airport hubs, which have been shown to play a crucial role for the diffusion of airborne diseases in many regions of the world (see for instance \cite{cacciapaglia2020better} and \cite{Mohammadi2023}).
To illustrate this, we plot a timeline of the initial spread of COVID-19 starting from February 24 until March 16, 2020, hence over a 4-week period. In each geographical unit, we identify the time of transition from a disease-free state to the initial exponential increase in the incidence numbers. In Figure~\ref{fig19} we show the progression of the disease within core clusters (left panels) and states (right panels): each geographical unit is colored in red on the week when the exponential increase is first detected, then it turns green from the week after. The difference between the two columns is telling. In the core cluster analysis, we see that the disease started in two clusters located in northern California and western Nevada. During the second week, there was a spread to nearby clusters in the west (Oregon, Seattle-area and Iowa) as well as to major airport hubs in the central and eastern part of the USA. We can easily identify isolated red clusters around Boston, New York, Washington, Chicago, Atlanta, Houston and Los Angeles in panel a2 of Figure~\ref{fig19}. During the third week, the disease reaches the remainder of the territory, except for areas in Washington/Montana states and Alaska, which run red during the fourth week.  The corresponding state-level analysis, shown in the right panels, features a similar overall pattern, however important details are missing or diluted. In particular, the importance of airports, which are indirectly highlighted in the core cluster analysis, is missing. Instead, the core cluster analysis confirms the results obtained, for instance, in \cite{cacciapaglia2020better}, where evidence was collected that airborne traffic was the principal culprit for the spread of COVID-19 from California to the rest of the country with an analysis of data aggregated at census division level. 

Furthermore, the use of the clustering algorithm allows to see specific features about the disease spread in local areas within some large states and also spanning across states. For instance, one can see that the disease starts effectively spreading in Texas near Houston and at the southern tip during the second week, while at state level one would conclude that Texas was one of the first affected states. These specific differences could inform policy, and could give decision makers and public health officials more tools to act on preventive measures very early on, rather than being guided by state-level views.

For visual confirmation of our results showing the importance of case importation by air, we plot the side-by-side panel a2 of Figure~\ref{fig19} and the enplanements map at the top 50 airports in the US, courtesy of the US Bureau of Transportation Statistics \href{https://www.bts.gov/enplanements-top-50-us-airports-2015}{\it https://www.bts.gov/enplanements-top-50-us-airports-2015} in Figure~\ref{hubs} below.
In the left panel we see the red spots corresponding to airline hubs, while in the right panel we see the flight density (represented by the size of each bubble) around the top 50 airports in US.
\begin{figure}
\centering
\includegraphics[scale=0.35]{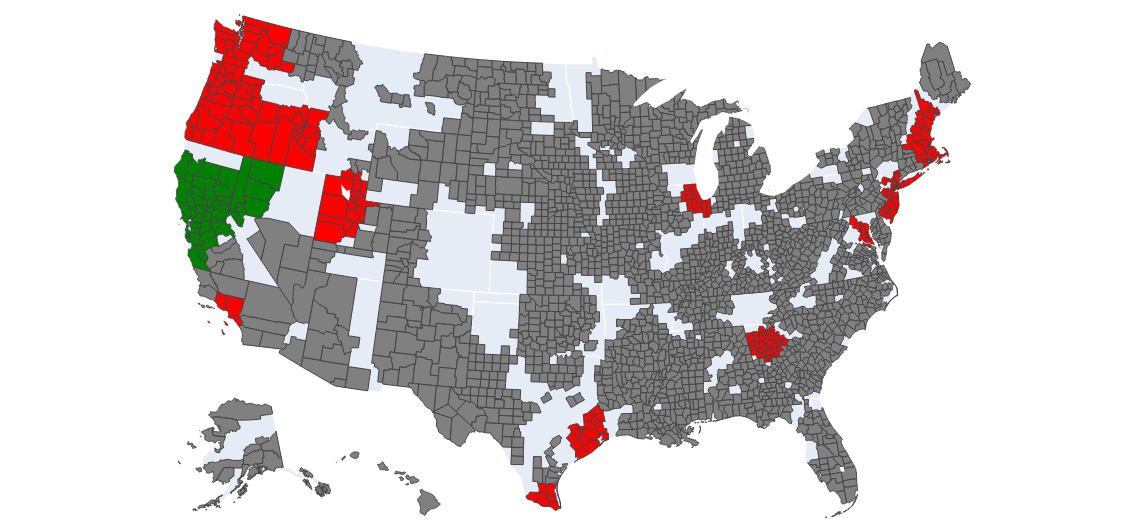}
\includegraphics[scale=0.1]{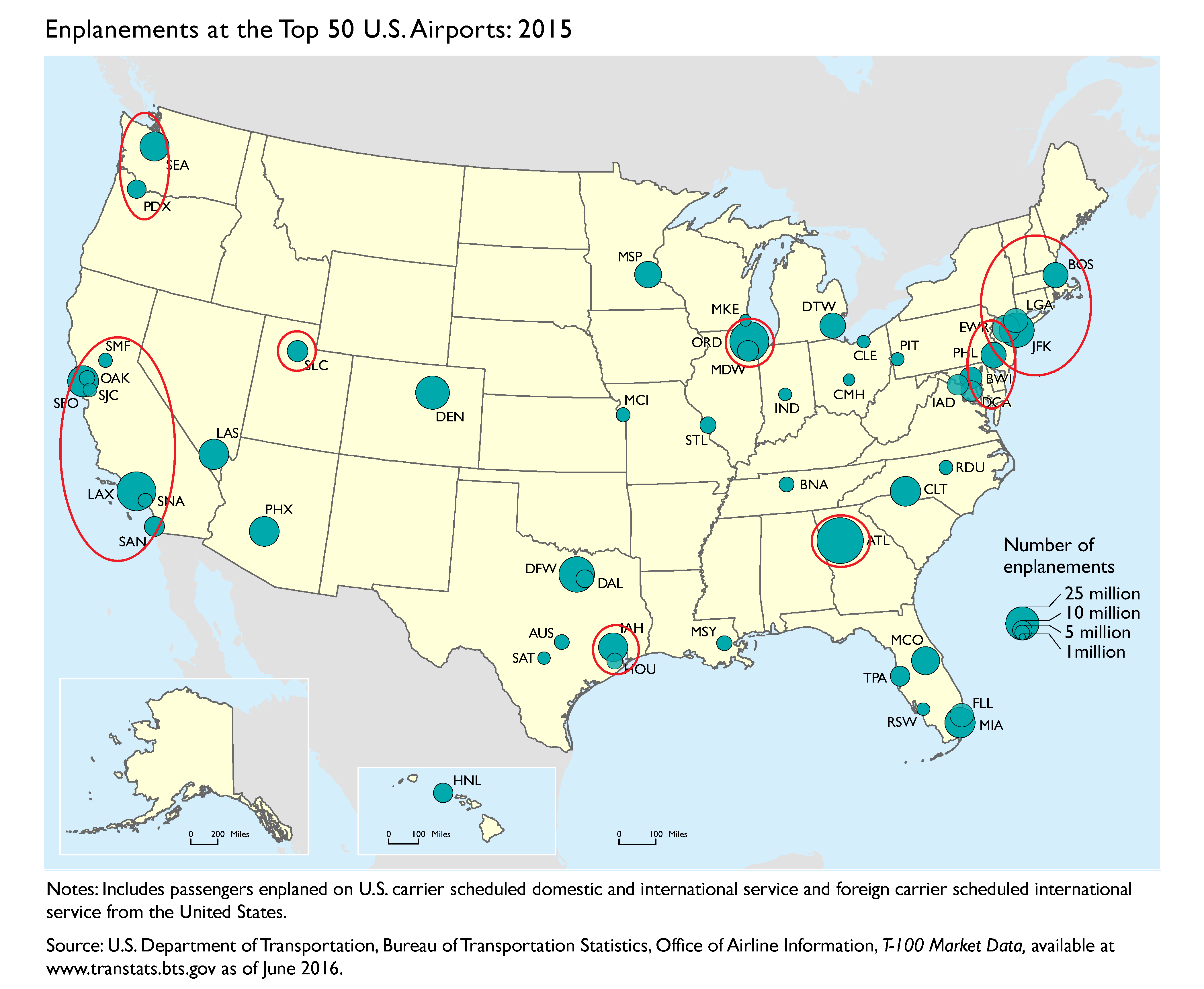}
\caption{To highlight the flight hubs in the red areas of the left panel (i.e., those with positive cases on the Week of March 1st, 2020), we marked the airports in those areas on the right panel map. There are some of the busiest (large bubbles correspond to high density of enplanements) in the US. }
\label{hubs}
\end{figure}

\subsection{Core Cluster vs. State epidemiology}

In order to capture the state-level versus the core cluster based view of the population and the concurrent disease spread, we extracted a few sample states with their corresponding clusters. We observed three possible situations: i) a state is essentially its own cluster without interactions with other clusters (e.g. Alaska and Maine); ii) a state contains several clusters (e.g., California and Florida), and iii) a cluster overlaps more than one state land mass. 
Case i) illustrates a trivial equivalence between the two methods, hence in this section we focus the analysis on examples of cases ii) and iii) above. 

\begin{figure}[H]
\begin{centering}
\includegraphics[scale=0.22]{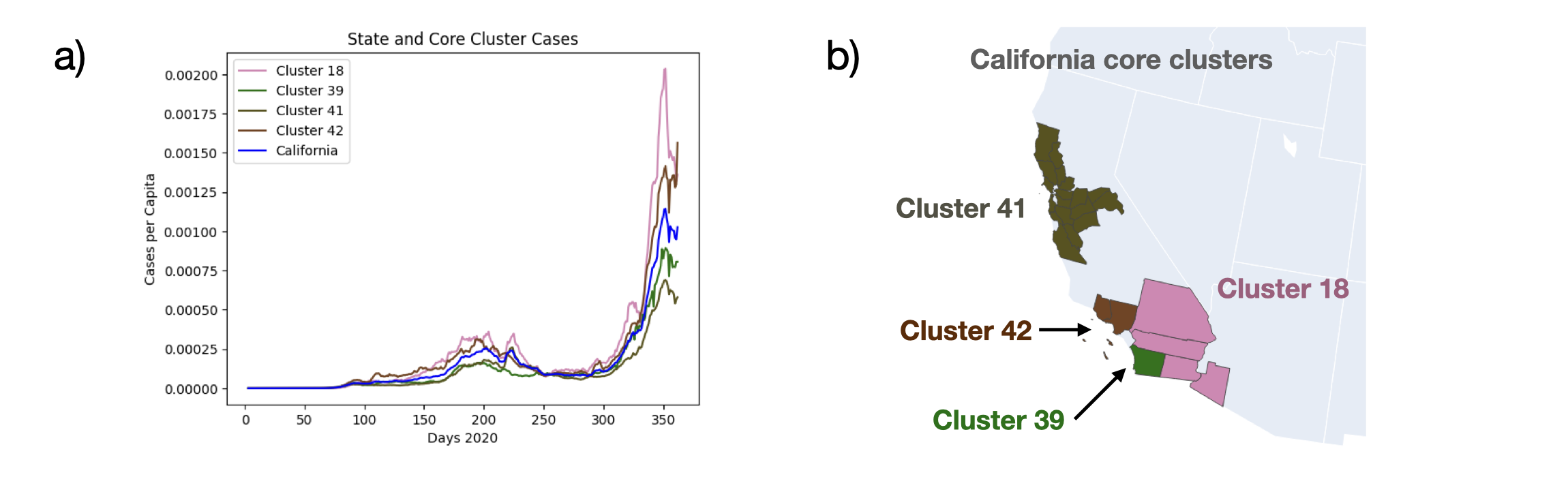}
\caption{Smoothed daily COVID-19 cases per capita for 2020 for California  as compared to the core clusters (a). In the right panel (b) we show the counties comprised in each core cluster.}
\label{fig24}
\end{centering}
\end{figure}

California is a clear example of case ii), see Figure \ref{fig24}b, as the majority of its population and territory is comprised within 4 core clusters: 41, 42, 39 and 18 (note that 18 also includes one county from Arizona). In Figure \ref{fig24}a we show the cases per capita, per day, smoothed via a 7 day rolling average, for California (in black) as compared to its four core clusters. Notably, not all the state cases are included, as some counties are removed from core clusters or are contained in out-of-state clusters. 
Thanks to the core cluster analysis, we can identify areas within California where the disease activity seems higher than elsewhere. For instance, clusters 18 and 42 had higher than state average cases per capita, thus it stands to reason to localize non-pharmaceutical interventions or preventive treatments in those zones. In Table \ref{tableCA} we list the counties comprised within clusters 18 and 42 and their population. Cluster 42 corresponds to the urban area of Los Angeles, and it saw an early rise of cases compared to the counties in nearby cluster 42. Instead, cluster 18 shows a larger incidence number at later stages in the pandemic: this makes sense, as the LA area has a much higher population density (so spread is very likely at rapid pace), and it was first detected as case positive by our analysis in Figure~\ref{fig19} very early, Week of March 1st 2020. Cluster 18 has a population density is a 10th of the LA-area and the initial spread happened a week later. It contains 3 Californian congressional districts (San Bernardino, Riverside and Imperial), with the first two districts known to be going back and forth from republican to democrat political representation. Yuma county, in Arizona, is similarly republican in presidential voting, with some democratic local representatives. Based on known correlations between willingness to adopt NPI measures and political leaning of US individuals (see \cite{hsiehchen2020political}), in a speculative way, our analysis seems to highlight the same argument: the higher than average incidence in cluster 18, though a lot sparser populated than 42, may be due to individual behavior, i.e., due to a higher presence of individuals disinclined to adopt NPI measures.  
 
\begin{table}[H]
\centering
\begin{tabular}{|c|c|c|c|}
\hline
Cluster 18 counties& Population & Cluster 42 counties & Population\\
\hline
    Yuma County  &  207829 & Los Angeles County & 10098052\\
    Imperial County &   180216 & Ventura County & 848112\\
    Riverside County & 2383286 &  & \\
    San Bernardino County & 2135413 & & \\
    \hline
    Population density & 132.77/sq mi & Population density & 1854.96/sq mi\\
    \hline
\end{tabular}
\caption{List of counties with their population enclosed within core clusters 18 and 42. The last row reports the average population density in the two clusters.}
\label{tableCA}
\end{table}

\begin{figure}[H]
\centering
\includegraphics[scale=0.21]{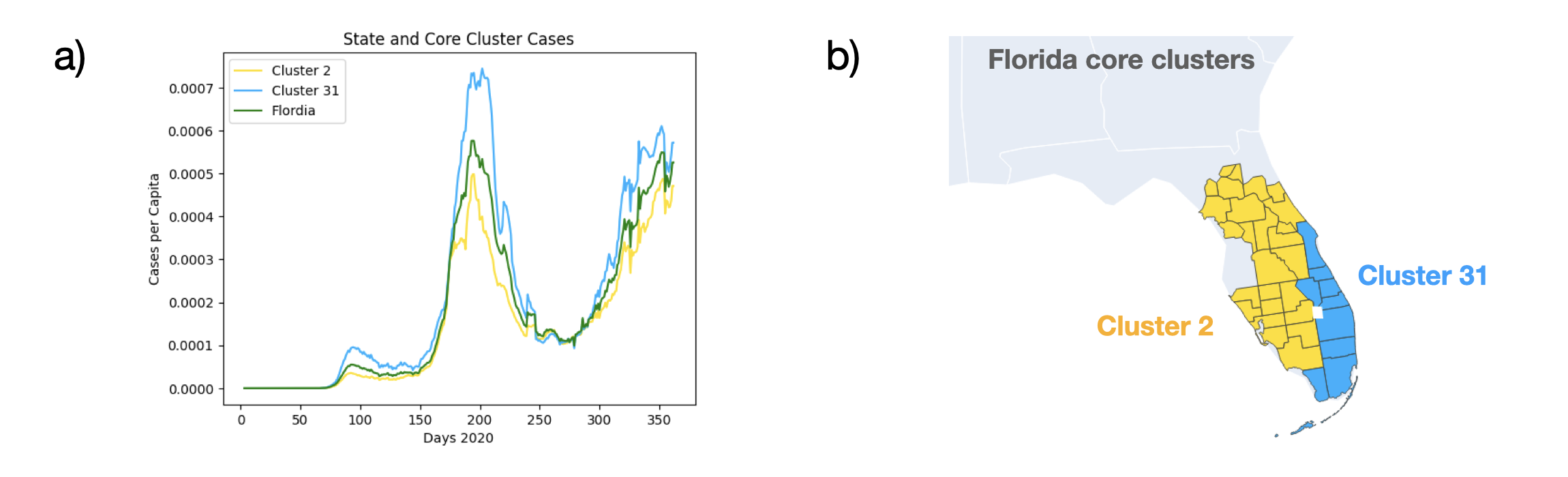}
\caption{Smoothed daily COVID-19 cases per capita for 2020 for Florida  as compared to the core clusters (a). In the right panel (b) we show the counties comprised in each core cluster.}
\label{Florida_Epi}
\end{figure}

We performed a similar analysis for the state of Florida, which comprises two enclosed clusters, as shown in Figure~\ref{Florida_Epi} (while counties in the north are joined with the neighboring states).
From panel \ref{Florida_Epi}a, we see that cluster 31 is disproportionately responsible for the spread during the 3 waves that Florida experienced in 2020, as compared to cluster 2. This effect could be explained by the higher population density in cluster 31, given by 780 people per square mile, as it also encloses Miami. Instead, cluster 2 has an average density of 318 people per square mile. Most likely, cluster 31 contains the counties of Monroe, Miami-Dade and Broward, the top 3 most tourist intensive, as reported at: \href{https://www.visitflorida.org/media/30679/florida-visitor-economic-impact-study.pdf}{\it https://www.visitflorida.org/media/30679/florida-visitor-economic-impact-study.pdf} and some of the nicest weather, thus providing ample opportunities for individuals to lower their risk perception of getting infected with Covid-19.

\begin{figure}[H]
\begin{centering}
\includegraphics[scale=0.21]{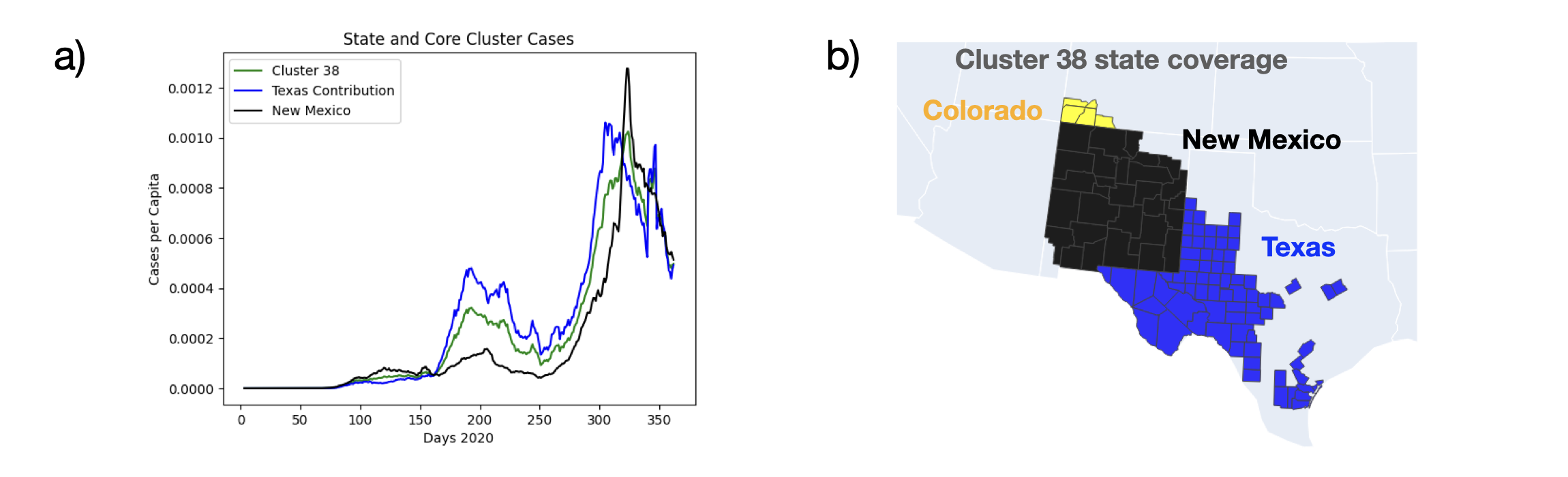}
\caption{Smoothed daily COVID-19 cases per capita for 2020 for core cluster 38 as compared to its counties from New Mexico and Texas (a). In the right panel (b) we show the cluster is subdivided among three states, New Mexico, Texas and Colorado.}
\label{NM_Epi_2}
\end{centering}
\end{figure}

To illustrate case iii) in this section, we took the example of the state of New Mexico: it is almost entirely covered by the much larger core cluster 38, which also includes a sizable part of Texas and a few counties in Colorado, as shown in Figure~\ref{NM_Epi_2}b.
In analogy to the analysis for cases ii) before, in Figure \ref{NM_Epi_2}a we show the cases per capita in the part of cluster 38 that overlaps with the states of New Mexico and Texas, versus the cases per capita in the whole cluster (in green). Interestingly, we see a rather different behavior in the two state portions of the cluster, due to the very different policies applied in the two states. Nevertheless, the interconnection among counties within the cluster, highlighted by our clustering algorithm, implies that disease could propagate from one side to the other far more easily that it could be expected by simply looking at the state boundaries.
Here the analysis implies that some coordination and collaboration on non-pharmaceutical interventions and preventive policies against the spread of disease would be beneficial to New Mexico, and it would help the state of Texas as well. Since Texas has had very little control over its disease spread during the pandemic, its policy had directly negatively influenced New Mexico.

\subsection{Analysis for Ontario, Canada}

Our algorithm is built generically, all that is required are regions that can be subdivided into $N$ sub-regions (such as counties or public health regions) and mobility data among sub-regions. To illustrate its versatility, we implemented it to other regions of the world, reporting here the case of the Ontario province in Canada. For the mobility information, in the case of Ontario, we switched from proprietary data to publicly available data. We used a 2016 work mobility survey as baseline for worker mobility and then adjusted it based on Google mobility index changes from baseline to obtain an interaction matrix. Moreover, we looked at Ontario as a collection of 34 public health regions, which are well-defined geographically. 

To achieve a daily contact rate between health units in the province of Ontario across 23 months (from February 2020 to December 2021), we combined data from two sources: mobility reports by Google \cite{googledata} and commuting flow data by the Government of Canada \cite{ commuterSC}. Both data sources are publicly available. 
Commuter flow data reveals \%25.16 of the employed population in Ontario works outside the census division where they live. This data source highlights that 11 census divisions out of 49  have more than \%40 of their workforce commuting daily outside the census division where they reside \cite{commuterSC}. By combining this data set with the Google Mobility reports, we can estimate the daily contact rates among the network nodes in Ontario.

During the COVID-19 pandemic, Google Mobility reports captured changes in movement over time compared to baseline (pre-lockdown) activity in different categories, such as retail/recreation, transit stations, and workplaces \cite{ googledata}. Google’s index data have been used in previous analyses \cite{mohammadi2022human, cot2021mining, fields2021covid}. Since Google split the mobility data into 51 regions in Ontario, corresponding to local municipalities, we had to combine several regions along their borders to obtain mobility data for 34 health units in this work (see \cite{smook2022adapting} for more details). The same borders were considered to calculate the commuter rate for health units, as commuter data was reported at the census division level.
Assume $m_{ij}$'s represent entries of the commuter flow matrix between health units in Ontario, where $i$ is the place of residence (POR) and $j$ is the place of work (POW). Thus, the contact rate between the health unit $i$, POR, and the health unit $j$, POW, on day $t$ is calculated as:
\begin{equation*}
    \text{contact}_{ij} = \frac{m_{ij}}{\text{Emp}_{i}} g_{i}^m(t), \quad  \text{where} \quad m=Work
\end{equation*}  
where $\text{Emp}_{i}$ is the employed population size of health unit $i$ and $g^m(t)$ represents the fluctuation percentage in the Google Work Index on day $t$ compared to the baseline Google Index.

\begin{figure}[H]
\begin{centering}
\includegraphics[scale=0.33]{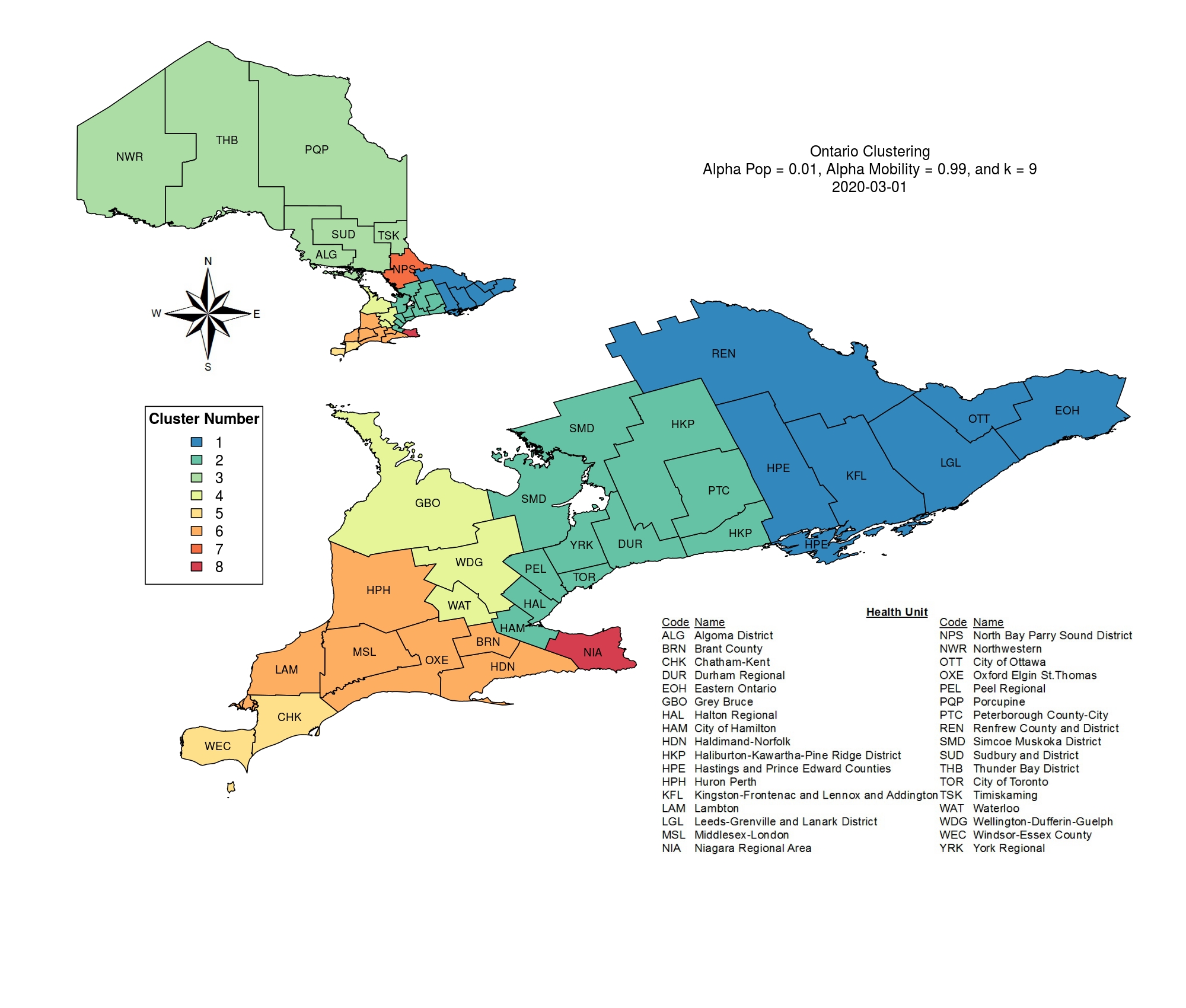}
\caption{Clustering of Ontario's health units with $\alpha_\text{Pop} = 0.01$ and $\alpha_\text{Int} = 0.99$.}
\label{fig28}
\end{centering}
\end{figure}

The result of the clustering algorithm is shown in Figure~\ref{fig28}, where we particularly focus on the southern region of Ontario, where most of the population is concentrated. This part is subdivided into 6 clusters (while two more are found in the northern part). Interestingly, the cluster structure roughly coincides with the administrative division of health units (\url{https://www.alphaweb.org/page/PHU}), where 5 regions are implemented: East, Central East, Central West, South West and Toronto. While East matches with the definition of cluster 1 (blue), which comprises Ottawa, the other regions are rearranged by our algorithm. Noticeably, Toronto is merged with Central East and part of Central West, following the intense commuter flow to the main city of the province from nearby regions. The other clusters clearly center over Niagara (8), Waterloo (4) and Windsor (5).
Hence, our cluster definition seems to better represent the demographic character of the province of Ontario as compared to the administrative division.

\section{Further Applications and Discussion} \label{sec5}

In this paper we formulated and studied a novel clustering algorithm based on mobility data among small geographical units, like counties. We applied it to the case of the counties in the USA across the first 6 months of the 2020 pandemic COVID-19, and we introduced the notion of {\it core clusters}. They are defined by counties that remain consistently into the same cluster over a long period, hence being insensitive to changes in mobility due to the implementation of local non-pharmaceutical measures.
The core clusters provide comparable geographical units characterized by a sub-population that is well-mixed by internal people's mobility. Hence, they offer an ideal basis to study the diffusion of an infectious disease within a large region. 
This new approach allowed us to capture the spatio-temporal spread of COVID-19 across USA and highlight features that are washed-out in a state-basis analysis. For instance, we found that the initial spread of COVID-19 in 2020 started from the north of California to a series of {\it hot-zones} that feature an airport hub. This result further confirms the relevance of airborne passenger transportation, as first highlighted in an earlier study using the nine census division as geographical units \cite{cacciapaglia2020better}. Furthermore, an analysis of the incidence number in core clusters showed that some states would need a higher granularity of localized measures to dampen disease spread while others showed the need for cooperation in dampening measures between neighboring states. The core clusters we identified in the USA could also be used as an efficient basis to predict the spread of an infectious disease across the country by use of a diffusion model, like for instance the eRG \cite{Cacciapaglia:2020mjf}. 

The work that was done to cluster Ontario was a simple example of the capabilities of our algorithm. Given any geographical area, the population distribution, and a form of mobility data, the algorithm is able to cluster virtually any region in order to discover well-connected sub-regions. This paper focuses on the usefulness of this algorithm on the COVID-19 pandemic, however no aspect of the algorithm is limited to the COVID-19 pandemic. As such, this algorithm is capable of being applied to a variety of problems that rely on population mobility. The fact that the algorithm is modular allows a user to be able to add additional terms onto the loss function. This allows for individuals to specifically tailor this algorithm to their problem. 
 
From a policy perspective, the versatility of applying our algorithm at any granularity level in a given large geographic area is of interest for governing bodies, who are concerned about differing sub-populations living and working there, and their connections with their neighbors. In the last section we pointed out specific regions with a disproportionate number of cases per capita relative to other regions, all within the same geographical/census area. Utilizing this information, areas which have experienced higher than state case numbers per capita during the pandemic can be extrapolated to have a higher probability of transmission of a pathogen such as SARS-COV-2, due to 
their internal well-mixing. Thus, they can be a target of resource allocation deployment, in order to reduce the impacts of a future pandemic/epidemic. On the other hand, cooperation in regional public policy would equally help, where two adjacent states, with well-mixing interstate populations, could try to deeply resources and preventive measures in a more coordinated fashion. Last but not least, we exemplify here several ways and uses of our study, however each aspect of our analyses can be developed more in-depth, depending on user needs.






\section*{Use of AI tools declaration}
The authors declare they have not used Artificial Intelligence (AI) tools in the creation of this article.

\section*{Acknowledgments}

All sources of funding for the study are disclosed below. Cojocaru, M.-G. acknowledges support for the work from the National Sciences and Engineering Research Council (NSERC), via a Discovery Grant 400684 and an Alliance Grant Option II (providing partial funding support for D. Lyver), and a Mathematics for Public Health Fields Institute grant (providing support for Z. Mohammadi). 
G.Cacciapaglia and C.Cot acknowledge partial support from the MITI project ``Événements rares'' of CNRS, project {\it SpikeRG}.
Cuebiq mobility data was purchased and made available to the authors by Sanofi.

\section*{Conflict of interest}

E.W.Thommes is an employee of Sanofi.

 \bibliography{bibliography}

\begin{thebibliography}{10}

\bibitem{anderson1979population}
R.~M. Anderson and R.~M. May.
\newblock Population biology of infectious diseases: Part i.
\newblock {\em Nature}, 280(5721):361--367, 1979.

\bibitem{arino2003}
J.~Arino and P.~van~den Driessche.
\newblock A multi-city epidemic model.
\newblock {\em Mathematical Population Studies}, 10(3):175--193, 2003.

\bibitem{arino2004}
J.~Arino and P.~Van Den~Driessche.
\newblock The basic reproduction number in a multi-city compartmental epidemic
  model.
\newblock In {\em Positive Systems: Proceedings of the First Multidisciplinary
  International Symposium on Positive Systems: Theory and Applications (POSTA
  2003), Rome, Italy, August 28--30, 2003}, pages 135--142. Springer, 2004.

\bibitem{cacciapaglia2020better}
G.~Cacciapaglia, C.~Cot, A.~S. Islind, M.~{\'O}skarsd{\'o}ttir, and F.~Sannino.
\newblock {Impact of US vaccination strategy on COVID-19 wave dynamics}.
\newblock {\em Scientific Reports}, 11:10960, 2021.

\bibitem{cacciapaglia2020second}
G.~Cacciapaglia, C.~Cot, and F.~Sannino.
\newblock {Second wave COVID-19 pandemics in Europe: A Temporal Playbook}.
\newblock {\em Sci Rep}, 10:15514, 2020.

\bibitem{Cacciapaglia:2020mjf}
G.~Cacciapaglia and F.~Sannino.
\newblock {Interplay of social distancing and border restrictions for pandemics
  (COVID-19) via the epidemic Renormalisation Group framework}.
\newblock {\em Sci Rep}, 10:15828, 5 2020.

\bibitem{Cauchy1847}
L.~Cauchy.
\newblock {Méthode générale pour la résolution des systèmes d'équations
  simultanées}.
\newblock {\em C.R. Acad. Sci. Paris}, 25:536--538, 1847.

\bibitem{chang2021}
S.~Chang, E.~Pierson, P.~W. Koh, J.~Gerardin, B.~Redbird, D.~Grusky, and
  J.~Leskovec.
\newblock Mobility network models of covid-19 explain inequities and inform
  reopening.
\newblock {\em Nature}, 589(7840):82--87, 2021.

\bibitem{cot2021mining}
C.~Cot, G.~Cacciapaglia, and F.~Sannino.
\newblock Mining google and apple mobility data: Temporal anatomy for covid-19
  social distancing.
\newblock {\em Scientific reports}, 11(1):4150, 2021.

\bibitem{fields2021age}
R.~Fields, L.~Humphrey, D.~Flynn-Primrose, Z.~Mohammadi, M.~Nahirniak,
  E.~Thommes, and M.~Cojocaru.
\newblock Age-stratified transmission model of covid-19 in ontario with human
  mobility during pandemic's first wave.
\newblock {\em Heliyon}, 7(9), 2021.

\bibitem{fields2021covid}
R.~Fields, L.~Humphrey, E.~W. Thommes, and M.~G. Cojocaru.
\newblock Covid-19 in ontario: Modelling the pandemic by age groups
  incorporating preventative rapid-testing.
\newblock In {\em Mathematics of Public Health: Proceedings of the Seminar on
  the Mathematical Modelling of COVID-19}, pages 67--83. Springer, 2021.

\bibitem{Fung:2001}
G.~Fung.
\newblock A comprehensive overview of basic clustering algorithms.
\newblock 2001.

\bibitem{golding2020tracking}
S.~A. Golding and R.~L. Winkler.
\newblock Tracking urbanization and exurbs: Migration across the rural--urban
  continuum, 1990--2016.
\newblock {\em Population research and policy review}, 39:835--859, 2020.

\bibitem{googledata}
{Google}.
\newblock Covid-19 community mobility reports, 2020.

\bibitem{Hevesi2006}
A.~G. Hevesi.
\newblock Outdated municipal structures.
\newblock {\em Office of the New York State Comptroller}, 2006.

\bibitem{hsiehchen2020political}
D.~Hsiehchen, M.~Espinoza, and P.~Slovic.
\newblock Political partisanship and mobility restriction during the covid-19
  pandemic.
\newblock {\em Public health}, 187:111--114, 2020.

\bibitem{kermack1927contribution}
W.~O. Kermack and A.~G. McKendrick.
\newblock A contribution to the mathematical theory of epidemics.
\newblock {\em Proceedings of the royal society of london. Series A, Containing
  papers of a mathematical and physical character}, 115(772):700--721, 1927.

\bibitem{ma2020estimating}
J.~Ma.
\newblock Estimating epidemic exponential growth rate and basic reproduction
  number.
\newblock {\em Infectious Disease Modelling}, 5:129--141, 2020.

\bibitem{Laszlo2007}
L.~Makra and Z.~Sümeghy.
\newblock Objective analysis and ranking of hungarian cities, with different
  classification techniques, part 2: Analysis.
\newblock pages 40--41, 01 2007.

\bibitem{Mohammadi2023}
Z.~Mohammadi, M.~Cojocaru, J.~Arino, and A.~Hurford.
\newblock Importation models for travel-related sars-cov-2 cases reported in
  newfoundland and labrador during the covid-19 pandemic.
\newblock {\em medRxiv}, 2023.

\bibitem{mohammadi2022human}
Z.~Mohammadi, M.~G. Cojocaru, and E.~W. Thommes.
\newblock Human behaviour, npi and mobility reduction effects on covid-19
  transmission in different countries of the world.
\newblock {\em BMC Public Health}, 22(1):1--19, 2022.

\bibitem{Rodriguez:2019}
M.~Z. Rodriguez, C.~H. Comin, D.~Casanova, O.~M. Bruno, D.~R. Amancio, L.~d.~F.
  Costa, and F.~A. Rodrigues.
\newblock Clustering algorithms: A comparative approach.
\newblock {\em PLOS ONE}, 14(1):1--34, 01 2019.

\bibitem{Ruder2016}
S.~Ruder.
\newblock An overview of gradient descent optimization algorithms.
\newblock {\em CoRR}, abs/1609.04747, 2016.

\bibitem{smook2022adapting}
S.~Smook.
\newblock {\em Adapting A Time-Dependent Vaccination Game to Mask Compliance}.
\newblock PhD thesis, University of Guelph, 2022.

\bibitem{commuterSC}
{Statistics Canada}.
\newblock {Commuting Flow from Geography of Residence to Geography of Work -
  Census Divisions: Sex (3) for the Employed Labour Force Aged 15 Years and
  Over Having a Usual Place of Work, in Private Households, 2016 Census - 25\%
  Sample Data}, 2018.
\newblock [Tables: 98-400-X2016391].

\bibitem{thommes2016absenteeism}
E.~Thommes, M.~Cojocaru, and S.~Athar.
\newblock Absenteeism impact on local economy during a pandemic via hybrid sir
  dynamics.
\newblock In {\em Dynamics of Disasters—Key Concepts, Models, Algorithms, and
  Insights: Kalamata, Greece, June--July 2015 2}, pages 309--328. Springer,
  2016.

\bibitem{Xu:2015}
D.~Xu and Y.~Tian.
\newblock A comprehensive survey of clustering algorithms.
\newblock {\em Annals of Data Science}, 02:165--193, 06 2015.

\end{thebibliography}
  \bibliographystyle{abbrv}

\section*{Supplementary}
\subsection{Comparison with existing clustering methods}
In order to address the problem of assessing how well the algorithm is clustering, a variation on the Stochastic Block model (SBM) is used. The SBM provides an ideal basis to ensure the accuracy of clustering algorithms [1]. An SBM forms a graph using the assumption that each node within a network belongs to a community and connects to other nodes within its own community with a probability $p$, and connects to other nodes not within its own community with probability $q$ [1]. Once this graph is formed, the goal is to be able to recover the communities based on the connections between nodes in a process called community detection [1]. Several types of recovery are possible based on the values of $p$ and $q$; exact recovery, partial recovery, and no recovery [1].

In order to achieve exact recovery using an SBM, the following inequality must hold [1]:
\begin{equation}
    (n(p-q))^2 > 2(n(p+q)),
\end{equation}
where $n$ is the number of nodes in the graph, and $p$ and $q$ are as defined previously. 

In order to use the SBM to test the algorithm, an example model was created such that the algorithm could be rigorously tested for performance. In order to create this test graph, three variables needed to be defined, $p$, $q$ and $n$. The definition of these variables remains the same as previously mentioned, but these variables will change throughout testing in order to better understand the algorithm's capabilities [1].

This investigation is aiding in establishing    optimal ranges for the weights of the loss function in equation (\ref{totalloss}). In order to determine the optimal values for $\alpha_{mobility}$ and $\alpha_{Pop}$ a grid search was performed. From the grid search, heat maps were created in order to view the performance of the algorithm. 

\begin{figure}[H]
\begin{centering}
\includegraphics[scale=.5]{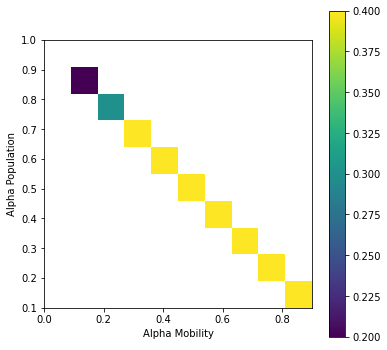}
\includegraphics[scale=.50]{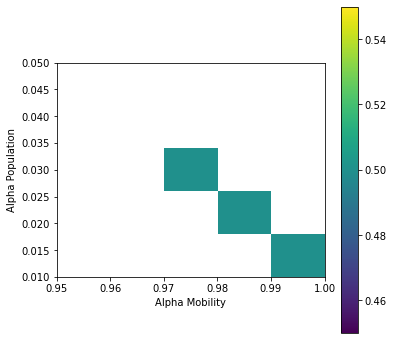}
\caption{Heat map for changing $\alpha_{Pop}$ and $\alpha_{mobility}$ values, where the color bar denotes the value of q, the number of communities $k = 5$, and the number of nodes $n = 500$.}
\label{fig6}
\end{centering}
\end{figure}

The performance in this case is defined by how close $q$ can be to $p$, such that the algorithm still performs exact recovery. Arbitrarily, the cluster number chosen was 5 and the number of nodes in the graph was 500; however, a range of values was being analyzed in order to understand the impact of manipulating each value. The initial test was done by looking over all possible $\alpha_{mobility}$ and $\alpha_{Pop}$, where $\alpha_{Pop}+\alpha_{mobility} = 1$ with a step size $0.1$. Each combination was tested on a set of $p$ and $q$ values, such that $p + q = 1$ and $p > q$. The initial test is visualized in Figure \ref{fig6}.

In Figure \ref{fig6} A, as the $\alpha_{mobility}$ value increases, the closer $q$ and $p$ are able to become before the algorithm is unable to exactly extract the communities. Continuing this process and reducing the step size allowed for a "zoomed-in" grid search for the region of best-fitting parameters. As well, the step size between the $p$ and $q$ values needed to shrink in order to find the boundary where the algorithm was unable to reconstruct the clusters for certain combinations of $\alpha$'s.

We continually saw an improvement of the boundary as $\alpha_{mobility}$ approaches $1$. The heat map for the final test run was used to determine the optimal value of the $\alpha's$ which can be seen in Figure \ref{fig6} B.

This results in an optimal range for $\alpha_{mobility} \in [0.97, 0.99]$ and $\alpha_{Pop} \in [0.03,0.01]$, which contains the estimated range that was assumed to provide the most accurate clusterization of the USA during previous testing.

Another interesting revelation that arose during this testing was the capability of this algorithm to be able to cluster past the bound of the Stochastic Block Model, SBM. Since the number of nodes was 500, the bound can be written as:
\begin{equation}
    (500(p-q))^2 > 2(500(1))
\end{equation}
Since $n = 500$, and $p + q = 1$, rearranging and solving for p generates the bound $p > 0.06 + q$ or $p > 0.53$. However, to obtain the results in Figure 5, the p-value is $.501$. This initially was a cause for concern; however, this is acceptable because of the additional information that is being provided to the system based on the population. This additional information is enough to allow the algorithm to detect communities past the theoretical bound. 
Since the number of nodes and the number of clusters were both arbitrarily chosen, the effects of altering these values on the algorithm's ability to extract the exact communities was examined. 

First, the effect of changing the number of clusters was investigated, and initially, it was expected that minimal effects would be seen by adjusting the number of communities. While in terms of the most accurate values for the $\alpha$ terms, this is true, the most accurate $\alpha_{Pop} \in [0.05, 0.01]$ and $\alpha_{mobility} \in [0.95, 0.99]$. However, there is effect on the accuracy of clustering when the number of communities increases.

\begin{figure}[H]
\begin{centering}
\includegraphics[scale=.5]{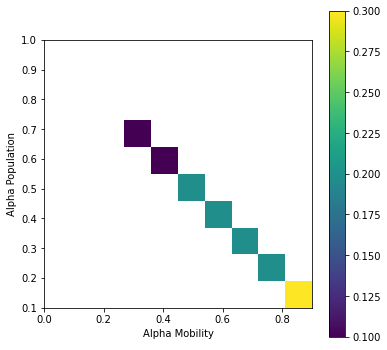}
\caption{Heat map for changing $\alpha_{Pop}$ and $\alpha_{mobility}$ values as $\alpha_{mobility}$ approaches 1, where the color bar denotes the value of q, the number of communities $k = 5$, and the number of nodes $n = 3000$.}
\label{fig8}
\end{centering}
\end{figure}

Next, the effect of changing the number of nodes while keeping the number of communities constant was analyzed, followed by the effect of both the number of communities and the number of nodes growing. This resulted once again in no change in terms of the optimal alpha values, as is demonstrated in Figure \ref{fig8} (below).

One may also note that even for the optimal $\alpha$ parameters, the closest that p and q can become until exact clustering is unable to be achieved is .7 and .3, respectively. A range of node values were analyzed up to $n = 3500$, but all resulted in a similar trend. The reason for stopping at $n=3500$ was due to the forseen applications of this algorithm. It is our goal to ensure that this algorithm is able to cluster the USA accurately, and the USA contains approximately, 3100 counties.

We also attempted to analyze the results of increasing both the number of communities and the number of nodes to be analogous with the problem that is trying to be solved in the USA. However, this yielded no notable results. 

The reason that some of these experiments yielded no notable results was because in increasing the number of communities and the number of nodes, the complexity of the problem increased. In order to understand the increase in complexity, it is critical to view what is being tested is whether a node has been placed in the correct community or if the node has been placed in the wrong community. This binary point of view allows the problem to be approached in a slightly different way. The number of connections between nodes in the same community grows by the following formula:
\begin{equation}
       p{\binom{\frac{n}{k}}{2}}
\end{equation}
For $n$ representing the number of nodes, $k$ representing the number of communities, and $p$ representing the probability a connection exists between two nodes with the same community. The number of connections between nodes outside its own community grows by the following formula:
\begin{equation}
    q(n - n/k)n
\end{equation}
Wherein $n$ and $k$ represent the same as in Formula 12 and q represents the probability that a node exists between two nodes that are not in the same community. 

Thus, as the number of nodes $n$ and the number of communities $k$ grows, there is a decrease in the performance of the algorithm. However, this decrease in performance is not seen in the clustering for the real-world application of the USA. The reason for this is due to the nature of human interaction within our collected data. The human interaction within our collected data causes a geological component to be introduced to the algorithm, as the flow metric that is being recorded is the actual travel between regions. This travel between regions is based on the geological location of each node, as for the most part, nodes that are very distant geographically have minimal to no travel between them. This in turn causes no large increase in the complexity of the problem, thus preventing the algorithm from failing. Importantly, this is only a hypothesis based on the properties of the real-world data versus the theoretical data that is generated from the SBM. This represents a topic that will require further research. An approach to this will likely involve similar techniques to that of the SBM, but a grid will be used in order to define distance within the network. Using this distance, it will be possible to only connect nodes that are in a neighbourhood around one another. Ideally, this distance metric will be capable to more closely capture the dynamics that are seen in real-world data.

\subsection{Accuracy}
\begin{figure}[H]
\begin{centering}
\includegraphics[scale=.5]{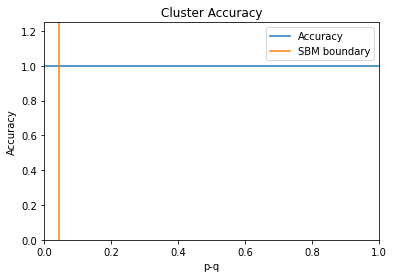}
\caption{Accuracy testing, for nodes = 1000 and k clusters = 5.}
\label{fig9}
\end{centering}
\end{figure}

Determining the accuracy of the model was done similarly to determining the best $\alpha$ values. Using the fixed $\alpha$ values $\alpha_{Pop} = 0.01$ and $\alpha_{mobility} = 0.99$, we ran the clustering algorithm on a set of $p$ and $q$ values in order to determine how well the algorithm clusters as it approaches the theoretical boundary 
Due to previous experiments to determine the optimal $\alpha$ values, it was assumed that for $k$ and $n$ values $k=5$ and $n=1000$ the algorithm would be able to cluster the nodes perfectly even well past the theoretical boundary. This is demonstrated in Figure \ref{fig9} (below).

The blue line in Figure \ref{fig9} represents the accuracy of clustering and shows that the algorithm is able to cluster the nodes into the correct communities past the theoretical boundary, which is represented by the orange horizontal line. 

\begin{figure}[H]
\begin{centering}
\includegraphics[scale=.5]{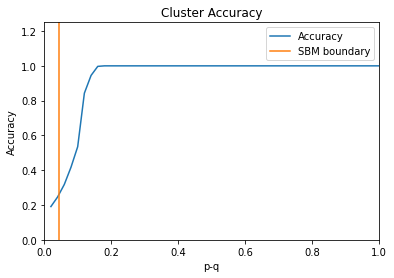}
\caption{Accuracy testing, for nodes = 1000 and k clusters = 10.}
\label{fig10}
\end{centering}
\end{figure}

By increasing the number of nodes, the number of communities, or both, the problem becomes more difficult and a drop in accuracy is expected. By increasing the number of communities to $k = 10$, this effect is shown in Figure \ref{fig10} (below).

Figure \ref{fig10} shows that, when clustering for a larger number of communities, the algorithm is no longer able to cluster past or even up to the theoretical boundary, as the problem has become too challenging. It is likely the accuracy of the algorithm would be affected less if a form of distance was introduced, as mentioned previously. Attempting to increase the number of nodes while keeping the number of communities constant had no notable effect on the accuracy of the algorithm, and thus it behaved very similarly to that of Figure \ref{fig9}.

\end{document}